\documentclass[letterpaper,12ppt]{article}
\usepackage[english]{babel}
\usepackage{amsthm}  % for math
\usepackage{amsmath} % for math
\usepackage{mathtools}% http://ctan.org/pkg/mathtools, for underbrace
\usepackage{amsfonts}
\usepackage{algpseudocode}
\usepackage{algorithm}
\usepackage{amssymb} % for more math
\usepackage{mathabx} % for even more math
\usepackage{graphicx}% for graphicx and figures
\usepackage{graphics}
\usepackage{fancyhdr}% for the footer
\usepackage{setspace}% for stuff like doublespacing
\usepackage{lineno}	 % for the linenumbers
\usepackage{lastpage}% for finding the last page (needed in the footer)
\usepackage{float}
\usepackage[utf8]{inputenc}
\usepackage{mathtools}
\usepackage{amssymb,amsmath,mathtools}
\usepackage{cite}
\usepackage{subcaption}
\usepackage{tabulary}
\usepackage[hyphens]{url}
\usepackage{caption}
\usepackage{comment}
\usepackage{subcaption}
\usepackage{lmodern} % Better font rendering
\usepackage[bottom]{footmisc}
\usepackage{xurl}

\usepackage
[
a4paper,% other options: a3paper, a5paper, etc
left=2.5cm,
right=2.5cm,
top=2.5cm,
bottom=2.5cm,
]
{geometry}

\usepackage{pgfplots}
\pgfplotsset{compat=1.11} 
\usepackage{graphicx}
\usepackage{xcolor}
\usepackage{svg}

\usetikzlibrary{shapes.geometric, arrows}

\newcommand{\mytitle}{Dynamic Hybrid Modeling: Incremental Identification and Model Predictive Control}
\newcommand{\myshorttitle}{Incremental Identification and MPC}
\newcommand{\myauthor}{Adrian Caspari$^{a,\ast}$, Thomas Bierweiler$^a$, Sarah Fadda$^b$, Daniel Labisch$^a$, Maarten Nauta$^b$, \\ Franzisko Wagner$^{a,c}$, Merle Warmbold$^{a,d}$, Constantinos C. Pantelides$^{b,e}$} % myauthor is important for the ^* !!!
\newcommand{\myauthorshort}{Caspari et al. 2025}
\author{\myauthor}
\usepackage[colorlinks,linkcolor=blue,citecolor=blue,urlcolor=blue,pdftitle={\mytitle},pdfauthor={Adrian Caspari}]{hyperref} 
\usepackage{cleveref}
\fancypagestyle{fancy}
{
	\fancyhead[R]{\myshorttitle}
	\fancyhead[L]{\myauthorshort}%\fancyhead[R]{\small{\textit{\the\day.\the\month.\the\year}}}
	\fancyfoot[L]{}
	\fancyfoot[R]{Page \thepage\ of \pageref*{LastPage}}
	\cfoot{}
}

\fancypagestyle{firststyle}
{ 

	\fancyhead[L]{}%\copyright \small{\textit{\myauthorshort}}}
	\fancyhead[C]{}
	\fancyhead[R]{}%Page \today \thepage\ of \pageref*{LastPage}}
	\fancyfoot[L]{\footnotesize
		$^{*}$Corresponding author: Dr.~Adrian Caspari, E-mail: \url{adrian.caspari@siemens.com}
	}
	\fancyfoot[C]{}
	\fancyfoot[R]{Page \thepage\ of \pageref*{LastPage}}
}

\addto\captionsenglish{%
}

\theoremstyle{remark} % to distinguish between different theorem styles (this is without italic)

\makeatletter
\let\@addpunct\@gobble
\g@addto@macro{\thm@space@setup}{\thm@headpunct{}} % point behind Def., Proof etc.
\makeatother

\renewenvironment{abstract}{\noindent\textbf{Abstract:}}{}

\theoremstyle{definition}

\begin{document}
\pagestyle{fancy}
\thispagestyle{firststyle}

\begin{center}
\begin{flushleft}\begin{Large}\begin{center}\textbf{\mytitle}\end{center}\end{Large} \end{flushleft}
\vspace{0.7cm}
\myauthor

\vspace{0.7cm}

\noindent $^a$ Siemens AG, Digital Industries, Process Automation, Strategy and Technology, \\ 76187 Karlsruhe, Germany \newline
\noindent $^b$ Siemens Industry Software Limited, Process Automation, Software, London W6 7HA, United Kingdom \newline
\noindent $^c$ Karlsruhe Institute of Technology, 76131 Karlsruhe, Germany \newline
\noindent $^d$ Clausthal University of Technology, 38678 Clausthal-Zellerfeld, Germany \newline
\noindent $^e$ Imperial College London, Department of Chemical Engineering, Sargent Center for Process Systems Engineering, London SW7 2AZ, United Kingdom

\end{center}

\noindent This is the author's accepted manuscript. The final published version can be accessed via the following DOI: \url{https://doi.org/10.1016/j.compchemeng.2025.109413}.

\vspace{0.7cm}
\hrule 
\vspace{0.7cm}
% !TeX encoding = UTF-8
% !TeX spellcheck = en_US
\begin{abstract} 
	Mathematical models are crucial for optimizing and controlling chemical processes, yet they often face significant limitations in terms of computational time, algorithm complexity, and development costs. Hybrid models, which combine mechanistic models with data-driven models (i.e. models derived via the application of machine learning to experimental data), have emerged as a promising solution to these challenges. However, the identification of dynamic hybrid models remains difficult due to the need to integrate data-driven models within mechanistic model structures.
	
	We present an incremental identification approach for dynamic hybrid models that decouples the mechanistic and data-driven components to overcome computational and conceptual difficulties. 
	Our methodology comprises four key steps: (1) regularized dynamic parameter estimation to determine optimal time profiles for flux variables, (2) correlation analysis to evaluate relationships between variables, (3) data-driven model identification using advanced machine learning techniques, and (4) hybrid model integration to combine the mechanistic and data-driven components. 
	This approach facilitates early evaluation of model structure suitability, accelerates the development of hybrid models, and allows for independent identification of data-driven components.
	
	Three case studies are presented to illustrate the robustness, reliability, and efficiency of our incremental approach in handling complex systems and scenarios with limited data.
\end{abstract}
\vspace{0.7cm}
\hrule

\vspace{0.7cm}


\vspace{0.7cm}

% !TeX encoding = UTF-8
% !TeX spellcheck = en_US
\section{Introduction}
\label{sec:introduction}

In chemical process optimization and control, mathematical modeling stands as a pivotal tool, offering profound insights into the behavior of complex dynamic systems, as required for instance in the context of model predictive control (MPC) \cite{Rawlings2017, Amrit.2013}. 
Historically, mechanistic models, grounded in fundamental principles and physical laws, have been extensively employed to describe chemical process behaviors \cite{Sansana2021, Bird1960, Caspari2019b, Caspari2019g}. 
However, such models are often limited by high computational demands, algorithmic complexity, high development costs, and difficulties in representing dynamic systems with incomplete understanding \cite{Caspari2019b, Caspari2019d, Schaefer2019, Schaefer2020a, Schaefer2019a, Schulze2020}. 
On the other hand, data-driven models usually lack transparency, generalization, and extrapolation capabilities but are often easier and cheaper to develop than mechanistic models \cite{Stosch.2014, Sansana2021, Shah2025}.
To address these challenges, hybrid models, which integrate mechanistic models with data-driven approaches (e.g., machine learning models such as artificial neural networks, ANNs), have emerged as a promising alternative. By combining the transparency and extrapolability of mechanistic models with the flexibility of data-driven methods, hybrid models offer improved accuracy and efficiency, particularly for dynamic systems requiring precise time-dependent behavior prediction and control \cite{Psichogios1992, Stosch.2014, Schaefer2020a, Sansana2021, Shah2025}.

The structure of hybrid models can broadly be classified into serial, parallel, and hybrid configurations \cite{Stosch.2014}. Among these, hybrid structures stand out for their versatility, as the data-driven and mechanistic components interact to directly influence system dynamics. However, this interdependence introduces significant challenges, as the two components cannot be handled independently during model identification. In contrast, serial and parallel structures allow independent identification of mechanistic and data-driven parts. 
Thus, the critical question arises as to how to efficiently identify hybrid models with hybrid structures.

Hybrid models with hybrid structures are typically identified using one of two approaches: simultaneous (or direct) and incremental identification \cite{Bardow2004, Sansana2021}. The simultaneous approach involves embedding the entire hybrid model, both mechanistic and data-driven parts, into a single parameter estimation problem. 
While conceptually straightforward, this method poses significant conceptual and computational challenges: the parameter estimation process involves numerous parameters, rendering it computationally intensive and unsuitable for standard machine learning training techniques. 
Moreover, diagnosing poor model performance is difficult, as issues may result from unsuitable mechanistic model parts, data-driven model parts, or parameter estimates. 
To overcome the limitations of simultaneous model identification approaches, incremental model identification approaches have been developed and applied \cite{Bardow2003, Bardow2004, Bardow2004a, Kahrs.2008, Kahrs2009, Asprion2018}. These methods decompose the identification process into sequential steps: mechanistic model calibration is performed independently, followed by data-driven model identification. This separation simplifies the identification process and enhances diagnostic clarity.
\begin{figure}[t]
	\centering
	\includegraphics[width=1\linewidth]{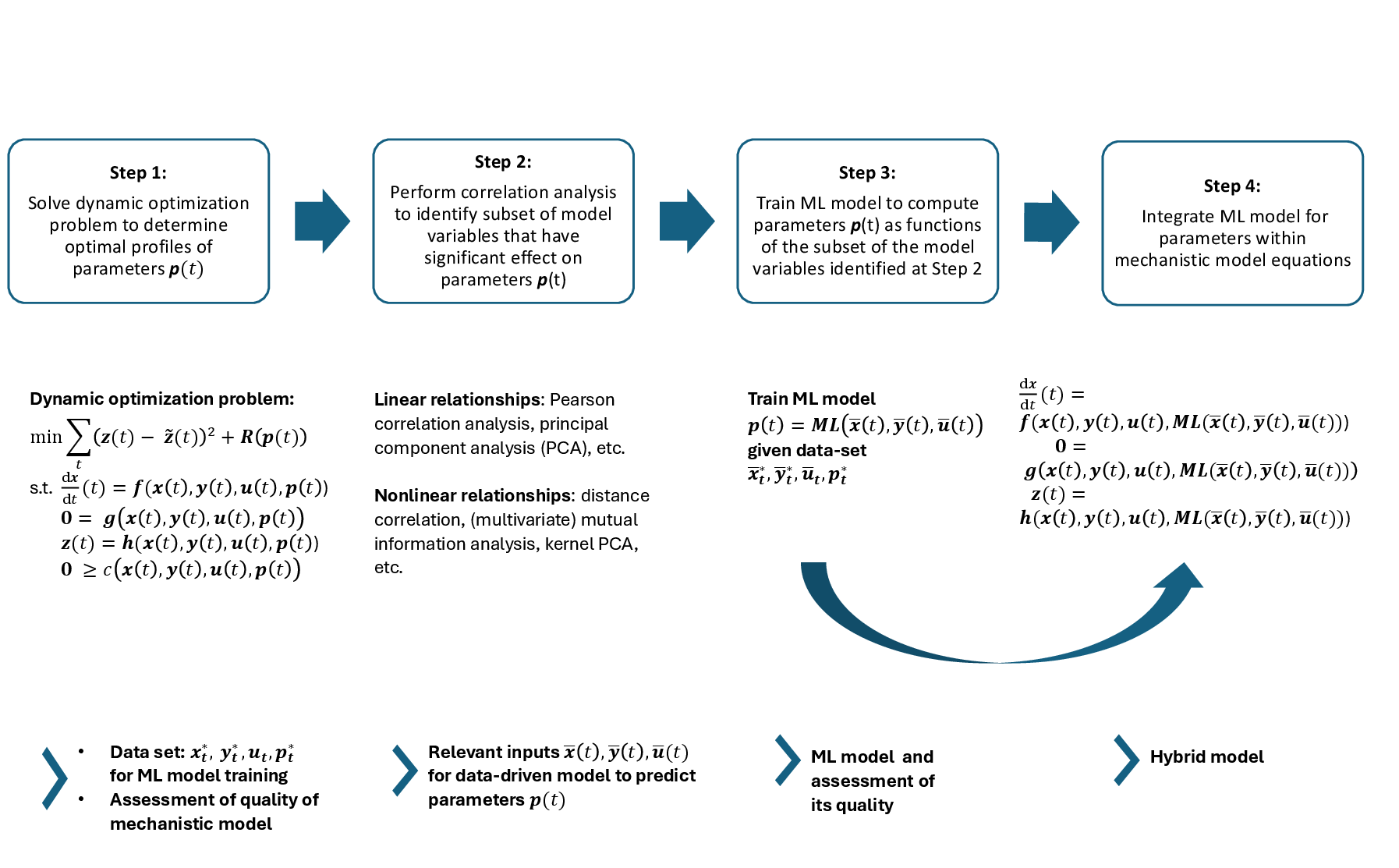}
	\caption{Overview of incremental hybrid model identification approach.}
	\label{fig:figsapproach1}
\end{figure}

Incremental identification approaches have been successfully developed and used to identify mechanistic models \cite{Bardow2003, Bardow2004, Bardow2004a}, steady-state hybrid models \cite{Kahrs.2008, Asprion2018}, and dynamic hybrid models \cite{Kahrs2009}. Bardow and Marquardt \cite{Bardow2003, Bardow2004, Bardow2004a} initially developed the incremental approach to identify mechanistic models. They investigated the incremental identification of transport coefficients in distributed systems, highlighting the potential of incremental approaches in characterizing complex transport phenomena \cite{Bardow2003}. 
Later, they focused on the identification of diffusive transport using an incremental approach, providing insights into the application of incremental methods in capturing transport dynamics \cite{Bardow2004a}. Additionally, they compared incremental and simultaneous identification methods for reaction kinetics, demonstrating the utility of incremental approaches in modeling chemical reaction kinetics within hybrid models \cite{Bardow2004}. Bardow and Marquardt \cite{Bardow2003, Bardow2004, Bardow2004a} did not consider hybrid models with general data-driven elements, i.e., elements without physical meaning. 
Instead, they selected the best fitting mechanistic models for the unknown model parts. Kahrs and Marquardt \cite{Kahrs.2008} generalized the approach of Bardow and Marquardt \cite{Bardow2003, Bardow2004, Bardow2004a}. They developed a general incremental identification approach that decomposes the identification problems into known and unknown parts and, hence, proposes a decomposition approach to efficiently solve the least-squares parameter estimation problem during model identification. 
They focused on stationary hybrid models. 
Later, Kahrs et al. \cite{Kahrs2009} applied the approach from \cite{Kahrs.2008} to dynamic hybrid models. 
They focused on dynamic hybrid models with a general regularization term in the dynamic parameter estimation problem and identified data-driven models for model parts with a physical meaning (e.g., reaction kinetics only). 
Asprion et al. \cite{Asprion2018} also applied an incremental identification approach to a stationary chemical process.

This paper introduces an incremental identification approach specifically designed for general dynamic hybrid models. 
To address a deficiency of previous approaches, we introduce a regularization term in the dynamic parameter estimation problem to prevent large discontinuities in the flux profiles determined at the first stage of the incremental identification. 
Such large discontinuities have no physical basis, and, consequently, affect the quality of the data-driven elements identified at the second stage of the incremental identification. 
This problem is particularly pronounced when the amount of available experimental data is limited, which is often the case in practical applications.
Furthermore, we consider generally unknown model parts with or without physical meaning.

We apply the new incremental identification approach in three case studies. 
The first case study illustrates the identification approach by focusing on a chemical reactor. 
The second case study focuses on dynamic bioreactor models and illustrates the development of a dynamic hybrid model with a limited amount of data. 
{The third case study focuses on a real-world research plant.}
In this case study, we identify a dynamic hybrid model which we embed as a controller model in a MPC framework to control the research plant.

The article is structured as follows. Section \ref{sec:methods} explains the incremental identification approach for dynamic hybrid models. Section \ref{sec:caseStudies} presents the case studies. Finally, we draw conclusions in Section \ref{sec:conclusions}.

% !TeX encoding = UTF-8
% !TeX spellcheck = en_US

\section{Dynamic Hybrid Modeling Approach}
\label{sec:methods}
% taken and heavily inspired from word document patent
We consider hybrid models modeled as index-1 differential-algebraic equation systems (DAEs) \cite{Ascher1998} of the following form:
\begin{subequations}
	\begin{align}
		\frac{\mathrm{d}\boldsymbol{x}}{\mathrm{d}t}(t) & = \boldsymbol{f}(\boldsymbol{x}(t),\boldsymbol{y}(t),\boldsymbol{u}(t),\boldsymbol{p}(t)), \\
		\boldsymbol{0} & = \boldsymbol{g}(\boldsymbol{x}(t),\boldsymbol{y}(t),\boldsymbol{u}(t),\boldsymbol{p}(t)), 
	\end{align}
	\label{eq:DAE}
\end{subequations}
where $\boldsymbol{x}(t) \in \mathbb{R}^{N_x}$ are the differential states, $\boldsymbol{y}(t) \in \mathbb{R}^{N_y}$ are the algebraic variables, $\boldsymbol{u}(t) \in \mathbb{R}^{N_u}$ are the inputs or manipulated variables (MVs), $\boldsymbol{f}: \mathcal{X} \rightarrow \mathbb{R}^{N_x}$ and $\boldsymbol{g}: \mathcal{X} \rightarrow \mathbb{R}^{N_y}$ define the differential and algebraic equations of the DAE, where $\mathcal{X} = \mathbb{R}^{N_x} \times \mathbb{R}^{N_y} \times \mathbb{R}^{N_u}  \times \mathbb{R}^{N_p}$.
Boldface symbols denote vector-valued variables, functions or matrices, while non-bold symbols represent scalar quantities.
%We consider the unknowns of the model as parameters $\boldsymbol{p}(t) \in \mathbb{R}^{N_p}$, i.e., these parameters represent those parts which are to be represented by data-driven models. The question is how to identify the hybrid model, i.e., how can the unknown model parts indicated by the parameters $\boldsymbol{p}(t)$ be represented. This question is answered by the incremental hybrid modeling approach we present in this work. 

Note that our approach is not limited to index-1 differential-algebraic equations (DAEs). 
We adopt this canonical form here because index-1 DAEs are commonly encountered in process control applications, and higher-index DAEs can typically be transformed into index-1 systems using established techniques, such as the algorithm proposed by Pantelides \cite{Pantelides1988}.

The model in Equation~\eqref{eq:DAE} is under-specified due to the inclusion of an additional set of unknown variables, $\boldsymbol{p}(t) \in \mathbb{R}^{N_p}$, which are difficult—or even impossible—to describe mechanistically. 
These variables may represent, for example, the rates of chemical reactions or fluxes associated with mass and heat transfer phenomena, particularly in cases where the underlying physics is not fully understood. 
However, they need not have a direct physical interpretation. 
In some cases, they may serve as correction or “error” terms applied to specific mechanistic expressions or even to entire equations{, e.g., using the following form:
	\[
	\frac{d\boldsymbol{x}}{dt}(t) = \boldsymbol{f}(\boldsymbol{x}(t), \boldsymbol{y}(t), \boldsymbol{u}(t)) + \boldsymbol{p}(t).
	\]
	Here, the parameters \( \boldsymbol{p}(t) \) serve as additive corrections to account for unmodeled dynamics or uncertain phenomena. This structure is particularly useful when detailed mechanistic knowledge is unavailable or when the unknowns lack a clear physical interpretation. Our incremental identification approach is flexible and supports both embedded and additive formulations. 
}

{
	In large-scale systems where the unknown profiles $\boldsymbol{p}(t)$ lack a clear physical interpretation, determining which differential or algebraic equations should include these parameters is closely tied to the concept of identifiability \cite{Isermann2011}. 
	Model structures with guaranteed system-theoretical properties can be determined using semi-infinite programming approaches, as in \cite{Caspari2020}.
	A systematic methodology for selecting the appropriate placement and number of unknown profiles is beyond the scope of this work.
	Instead, we assume that the structure of the hybrid model \eqref{eq:DAE} is already determined by the model developer.
}

The incremental hybrid modeling approach proposed in this work leverages machine learning applied to experimental data to represent the unknown variables $\boldsymbol{p}(t)$ as functions of other system variables—namely, the differential states $\boldsymbol{x}(t)$, algebraic variables $\boldsymbol{y}(t)$, and manipulated variables $\boldsymbol{u}(t)$. 
Our approach is structured into the four steps:
\begin{enumerate} 
	\item Regularized dynamic parameter estimation, 
	\item Correlation analysis, 
	\item Data-driven model identification, 
	\item Hybrid model integration. 
\end{enumerate}

Each of these steps is detailed in the following sections. 
An overview of the complete approach is illustrated in Fig.~\ref{fig:figsapproach1}.

\subsection{Regularized Dynamic Parameter Estimation}
\label{sec:methods:step1}
We assume the availability of one or more experimental datasets, each obtained by applying a known variation of the manipulated variables (MVs) $\boldsymbol{u}(t)$. Each dataset contains measured values $\boldsymbol{\tilde{z}}_j \in \mathbb{R}^{N_z}$ for all $j \in \{1, \dots, N_\mathrm{meas}\}$, recorded at $N_\mathrm{meas}$ distinct measurement time points $t_j \in \mathcal{T}^\mathrm{meas}$, where $\mathcal{T}^\mathrm{meas} = \{ t^\mathrm{meas}_1, \dots, t^\mathrm{meas}_{N_\mathrm{meas}} \}$ and $t^\mathrm{meas}_{j+1} > t^\mathrm{meas}_j$ for all $j \in \{1, \dots, N_\mathrm{meas} - 1\}$.

In the first step of our approach, we determine the optimal time profiles of the parameters $\boldsymbol{p}(t)$ for each experimental dataset such that the predictions of the model~\eqref{eq:DAE} best match the corresponding experimental measurements. This is formulated as the following dynamic optimization problem, solved independently for each dataset:
\begin{subequations}
	\begin{align}
		\underset{\boldsymbol{x},\boldsymbol{y},\boldsymbol{p}, \boldsymbol{z}, \boldsymbol{x}_0}{\text{min}} \sum_{\substack{j \in \\ \{1,...,N_\mathrm{meas}\}}} & \big( \boldsymbol{z}_j - \boldsymbol{\tilde{z}}_j \big)^\mathrm{T} \cdot \boldsymbol{W}_{j}^\mathrm{z} \cdot \big( \boldsymbol{z}_j - \boldsymbol{\tilde{z}}_j \big) + R(\boldsymbol{p}(t)) \label{eq:parameter_estimation:obj} \\
		s.t. \frac{\mathrm{d}\boldsymbol{x}}{\mathrm{d}t}(t) & = \boldsymbol{f}(\boldsymbol{x}(t),\boldsymbol{y}(t),\boldsymbol{u}(t),\boldsymbol{p}(t)), \ \forall t \in \mathcal{T} \\
		\boldsymbol{0} & = \boldsymbol{g}(\boldsymbol{x}(t),\boldsymbol{y}(t),\boldsymbol{u}(t),\boldsymbol{p}(t)), \ \forall t \in \mathcal{T} \\
		\boldsymbol{x}(t^\mathrm{meas}_1) & = \boldsymbol{x}_0 \\
		\boldsymbol{z}_j & = \boldsymbol{h}(\boldsymbol{x}(t_j),\boldsymbol{y}(t_j)), \ \forall t_j \in \mathcal{T}^\mathrm{meas} \\
		\boldsymbol{0} & \geq \boldsymbol{c}(\boldsymbol{x}(t),\boldsymbol{y}(t),\boldsymbol{u}(t),\boldsymbol{p}(t)), \ \forall t \in \bar{\mathcal{T}}, \label{eq:parameter_estimation:eqf}
	\end{align}
	\label{eq:parameter_estimation}
\end{subequations}
with the time horizon $\mathcal{T} = [ t^\mathrm{meas}_1, t^\mathrm{meas}_{N_\mathrm{meas}} ]$, the initial states $\boldsymbol{x}_0 \in \mathbb{R}^{N_x}$, the output variables $\boldsymbol{z}_j \in \mathbb{R}^{N_z}$ at the distinct time points $t_j \in \mathcal{T}^\mathrm{meas}$, the output equations $\boldsymbol{h}: \mathcal{X} \rightarrow \mathbb{R}^{N_z}$, {inequality} constraints $\boldsymbol{c}: \mathcal{X} \rightarrow \mathbb{R}^{N_g}$, that need to be satisfied over a subset $\bar{\mathcal{T}} \subseteq \mathcal{T}$ of the time domain $\mathcal{T}$, and the regularization function $R: \mathbb{R}^{N_p} \rightarrow \mathbb{R}$. 
$\boldsymbol{W}_j^z \in \mathbb{R}^{N_z \times N_z}$, $j \in \{1,...,N_\mathrm{meas}\}$ is the covariance matrix of the measurements $\tilde{\boldsymbol{z}}_j$ at time point $t_j$.

The MVs $\boldsymbol{u}(t)$ are known time profiles and are treated as fixed inputs during the optimization. 
In contrast, the parameter profiles $\boldsymbol{p}(t)$ are unknown and must be estimated. We assume that $\boldsymbol{p}(t)$ follows a piecewise constant profile. 
To this end, we define a discretization grid $t_k^{\mathrm{disc}}, \, k = 1, \dots, N_{\mathrm{disc}}$ with $t_1^{\mathrm{disc}} = t_1^{\mathrm{meas}}$ and $t_{N_{\mathrm{disc}}}^{\mathrm{disc}} = t_{N_{\mathrm{meas}}}^{\mathrm{meas}}$. 
The discrete parameter values are denoted by $\boldsymbol{p}_k$ for $k = 1, \dots, N_{\mathrm{disc}}$, such that $\boldsymbol{p}(t) = \boldsymbol{p}_k$ for $t \in [t_{k-1}^{\mathrm{disc}}, t_k^{\mathrm{disc}})$, $k = 2, \dots, N_{\mathrm{disc}}$.

The first part of the objective function \eqref{eq:parameter_estimation:obj} is a weighted least-squares objective penalizing the deviation of the model output variable values from their corresponding measurement values. 
The optimization may also include the constraints \eqref{eq:parameter_estimation:eqf} designed to ensure that the optimal choice of the parameters $\boldsymbol{p}_k$ does not result in physically unrealistic solutions. 
In practice, such constraints usually represent upper or lower bounds on either the values or the rate of variation of individual model variables. 
However, more complex relations (e.g. limitations on the relative values of pairs of variables) can also be accommodated.

The dynamic optimization problem \eqref{eq:parameter_estimation} may be ill-behaved, with multiple time profiles $\boldsymbol{p}(t)$ resulting in the same or very similar fits of predictions to experimental data. 
This is especially likely when only limited experimental data is available. 
In practice, this situation often results in the optimal solution exhibiting large jumps in the values of $\boldsymbol{p}(t)$ from one discretization interval to the next. 
From the point of view of the hybrid model identification, this is problematic: as we shall see in Section \ref{sec:methods:step3}, the parameters $\boldsymbol{p}(t)$ will ultimately be expressed, e.g., via machine learning models, as continuous functions (e.g. ANNs) of subsets of the model variables $\boldsymbol{x}(t)$, $\boldsymbol{y}(t)$, $\boldsymbol{u}(t)$ which typically represent physical quantities that may not exhibit such discontinuous behavior.
Overall, the discrepancy between the smoothness of $\boldsymbol{p}(t)$ and that of $\boldsymbol{x}(t)$, $\boldsymbol{y}(t)$, $\boldsymbol{u}(t)$ may significantly affect the quality of the data-driven model that can be identified based on the results of the dynamic optimization problem. 
%{Discontinuities in the estimated parameter profiles \( \boldsymbol{p}(t) \) arise since the variable profiles in the optimization problem \eqref{eq:parameter_estimation} are discretized. As a result, discontinuities are inherent to the formulation. When experimental data is limited or noisy, the optimization problem may become ill-posed, allowing multiple parameter trajectories to yield similar fits to the data. This can lead to abrupt changes (large discontinuities) in \( \boldsymbol{p}(t) \) between adjacent time intervals. To mitigate these effects, we introduce a regularization term that penalizes large changes in the parameter profiles \( \boldsymbol{p}(t) \) across time intervals. This promotes smoother and more physically plausible profiles, especially in scenarios with sparse data. }
In view of the above, the second part of the objective function \eqref{eq:parameter_estimation:obj} is a weighted regularization term that penalizes changes in the values of the parameters $\boldsymbol{p}(t)$ between successive discretization intervals:
\begin{equation}
	R(\boldsymbol{p}(t)) = \sum_{\substack{k \in \\ (1,...,N_\mathrm{disc}-1)}} \big( \boldsymbol{p}(t_{k+1}^\mathrm{disc}) - \boldsymbol{p}(t_k^\mathrm{disc}) \big)^\mathrm{T} \cdot \boldsymbol{W}^\mathrm{R}_{k} \cdot \big( \boldsymbol{p}(t_{k+1}^\mathrm{disc}) - \boldsymbol{p}(t_k^\mathrm{disc}) \big)
	\label{eq:regularization}
\end{equation}
The values of the regularization parameters $\boldsymbol{W}_k^R$ need to be selected carefully to avoid over-penalizing changes in $\boldsymbol{p}(t)$ in time periods where there may, indeed, occur variations in the state variables $\boldsymbol{x}(t)$, $\boldsymbol{y}(t)$, e.g. caused by large discontinuous changes in the values of the MVs $\boldsymbol{u}(t)$.
{Section~\ref{sec:caseStudies:chemicalReactor:regularization} provides a comparative analysis illustrating the impact of the regularization on the smoothness and quality of the estimated profiles.}

Solving the dynamic optimization problem \eqref{eq:parameter_estimation} yields the optimal time-varying profiles of the parameters $\boldsymbol{p^*}(t)$, differential states $\boldsymbol{x^*}(t)$, and algebraic variables $\boldsymbol{y^*}(t)$, given the profiles of the MVs $\boldsymbol{u}(t)$.
In practice, we solve the optimization problem \eqref{eq:parameter_estimation} using either full discretization or single shooting, cf. \cite{Biegler.2010, Caspari2019a}.
In the former case, we employ the same discretized time grid $T^{disc}$ as the one already introduced for the parameters $\boldsymbol{p}(t)$.

Once the solution of the dynamic parameter estimation optimization problem \eqref{eq:parameter_estimation} is obtained, it is possible to assess the performance of the mechanistic model part. 
In particular, if the optimal profiles of the variables $\boldsymbol{p}^\ast(t)$ fail to result in a sufficiently good match of model predictions $\boldsymbol{z}(t_j)$ with experimental measurements $\tilde{\boldsymbol{z}}_j$, the mechanistic model has to be adjusted. 
It should be noted that the fit depends on the weightings of the weighted square sum $\boldsymbol{W}^\mathrm{z}_j$ and of the regularization term $\boldsymbol{W}^\mathrm{R}_k$. 
While larger weights for the regularization term lead to a smoother profile for the parameters, they might also have a negative impact on the model fit.

It should be noted that the above approach is not restricted to piecewise constant profiles for the parameters $\boldsymbol{p}(t)$, and could be applied to other forms of discretization with an appropriate adaptation of the regularization function.  
For example, for piecewise linear continuous profiles, the regularization term would penalize changes in the \textit{slopes} of $\boldsymbol{p}(t)$ between successive discretization intervals.

\subsection{Correlation Analysis}
\label{sec:methods:step2}
The solution of the dynamic optimization problem (cf. Section \ref{sec:methods:step1}) for each experimental data set establishes the values of the parameters $\boldsymbol{p}^\ast(t_k^{disc})$ at the discrete time points $t_k^{disc} \in \mathcal{T}^\mathrm{disc}$, and also the values of the differential variables $\boldsymbol{x}^\ast$, algebraic variables $\boldsymbol{y}^\ast$, and MVs $\boldsymbol{x}$ at these same points.  
Overall, we denote these values as a set of triplets $({x}^\ast_k, {y}^\ast_k, {u}_k, {p}^\ast_k)$, $\forall k=1,..,N_{disc}$.

Note that due to the index-1 property of model \eqref{eq:DAE}, the values of the algebraic variables $\boldsymbol{y}(t)$ at any time $t$ are fully determined by the corresponding values of the differential states $\boldsymbol{x}(t)$ and the MVs $\boldsymbol{u}(t)$.
It is, therefore, not strictly necessary to use the algebraic variables $\boldsymbol{y}(t)$ later as an input to the data-driven models for the parameters $\boldsymbol{p}(t)$.
Nevertheless, it can be beneficial to do so as this could reduce the number of inputs to the data-driven models and, thus, reduces the risk of extrapolation.
%Moreover, if there are more than one experimental data set, the dynamic optimization problem can be solved independently for each one of them to determine the corresponding set of triplets   $({x}^\ast_k, {y}^\ast_k, {u}_k, {p}^\ast_k)$, $\forall k=1,..,N_{disc}$ for each data set. 

We can now concatenate the set of triplets arising from all the experimental datasets into a single table which can then be used for data analysis independently of the mechanistic model.

As a first step, we perform a correlation analysis to identify the subset of the variables $\boldsymbol{x}(t)$, $\boldsymbol{y}(t)$, $\boldsymbol{u}(t)$ which have a significant effect on the parameters $\boldsymbol{p}(t)$ and should, therefore, be used as inputs to a machine-learned model used to predict $\boldsymbol{p}(t)$. 
While various sophisticated methods (e.g. transfer entropy analysis, cf. \cite{Schreiber2000, 2023_Caspari-AlarmCauseEffect}) exist for calculating correlations from data or information flow we employ the straightforward Pearson correlation analysis to identify which variables should be used as inputs to predict the values of the parameters $\boldsymbol{p^*}(t)$. 
Specifically, if the absolute value of the correlation coefficient $c_{x,p}$ between a parameter $p$ and a variable $x$ exceeds a specified threshold $\tau$, i.e., $| c_{x,p} | \geq \tau$, we consider the parameter $p$ to be significantly correlated with the variable $x$ and include $x$ as an input for predicting $p$.

\subsection{Data-Driven Model Identification}
\label{sec:methods:step3}
The correlation analysis described in Section \ref{sec:methods:step2} identifies the subsets $\bar{\boldsymbol{x}}$,  $\bar{\boldsymbol{y}}$, and $\bar{\boldsymbol{u}}$ of, respectively, the differential states ${\boldsymbol{x}}$, algebraic variables ${\boldsymbol{y}}$, and/or the MVs ${\boldsymbol{u}}$ that have a significant effect on the parameters $\boldsymbol{p}$. 
Using the corresponding columns in the data table established by the solution of the dynamic optimization problem(s) (cf. Section \ref{sec:methods:step1}), we can now use machine learning to train a model to establish a mapping $\boldsymbol{p}=\boldsymbol{ML}(\bar{\boldsymbol{x}},\bar{\boldsymbol{y}}, \bar{\boldsymbol{u}})$.
Since this training is not embedded within any dynamical structure, standard libraries and approaches for artificial neural network (ANN) training can be utilized.
If necessary, hyper-parameter optimization can be performed using standard libraries such as TensorFlow\footnote{TensorFlow, \url{https://www.tensorflow.org/}, accessed: 14.04.2025} to enhance the performance of the data-driven model.

\subsection{Hybrid Model Integration}
In the last step, the trained data-driven models $\boldsymbol{ML}(\bar{\boldsymbol{x}}(t), \bar{\boldsymbol{y}}(t), \bar{\boldsymbol{u}}(t))$ are used to replace the parameters $\boldsymbol{p}(t)$ in Equations \eqref{eq:DAE} to obtain the final hybrid model. 
This yields the following hybrid model:
\begin{subequations}
	\begin{align}
		\frac{\mathrm{d} \boldsymbol{x}}{\mathrm{d} t}(t) & = \boldsymbol{f}(\boldsymbol{x}(t),\boldsymbol{y}(t),\boldsymbol{u}(t), \boldsymbol{ML}(\bar{\boldsymbol{x}}(t), \bar{\boldsymbol{y}}(t), \bar{\boldsymbol{u}}(t))), \\
		\boldsymbol{0} & = \boldsymbol{g}(\boldsymbol{x}(t),\boldsymbol{y}(t),\boldsymbol{u}(t), \boldsymbol{ML}(\bar{\boldsymbol{x}}(t), \bar{\boldsymbol{y}}(t), \bar{\boldsymbol{u}}(t))).
	\end{align}
	\label{hybrid_model_full_formulation}
\end{subequations}

As a final check on the quality of the identified hybrid model \eqref{hybrid_model_full_formulation}, we can use it to simulate each of the original experimental datasets, and compute the corresponding value of the weighted deviation between model prediction and experimental measurements.  
We note that these deviations may be different (not necessarily worse) than those originally determined via the optimal solutions of the dynamic optimization problem \eqref{eq:parameter_estimation}. 
Such differences may arise because: 
{The} data-driven model derived at Step 3 (cf. Section \ref{sec:methods:step3}) may not be a sufficiently accurate representation of the dependence of $p$ on the other system variables.
{Moreover,} at Step 1 (cf. Section \ref{sec:methods:step1}), the parameter profiles $\boldsymbol{p}(t)$ were approximated as piecewise constant functions of time, but now they are treated as continuous functions of the other system variables.
In principle, a fine-tuning of the hybrid model could be performed at this stage  by solving a dynamic optimization problem incorporating multiple instances of the hybrid model \eqref{hybrid_model_full_formulation}, one for each experimental dataset. 
The parameters of the data-driven model (e.g. ANN weights and biases) would be treated as optimization decision variables.  
However, the solution of such a problem is likely to be very computationally expensive for anything other than simple models.

{A final confirmation of the hybrid model’s quality can be performed via standard techniques such as $k$-fold cross validation. Ultimately, the quality of the hybrid model will depend on (a) the quality of the underlying mechanistic model, (b) the way in which the data-driven elements are introduced within this mechanistic model, and (c) the quality and amount of the available experimental data.}

{The major advantages of {the} methodology described in this Section \ref{sec:methods} result from the identification being carried out in two stages.}
Step 1 (cf. Section \ref{sec:methods:step1}) determines time profiles for those quantities $\boldsymbol{p}(t)$ that are difficult to characterize mechanistically. 
Step 3 (cf. Section \ref{sec:methods:step3}) then establishes the relation between these quantities and the other system variables. 
This separation allows each step to be evaluated independently. 
If the measured and estimated states from Step 1 diverge significantly, the model structure from Step 1 must be revised as any such mismatch is independent of the data-driven structure used later. 
At Step 3, the data-driven model is determined entirely from a data table established at Step 1 without the need to take any further account of the original mechanistic model \eqref{eq:DAE}. 
This greatly facilitates hyper-parameter optimization and training using standard libraries and approaches like TensorFlow, something which is problematic for any simultaneous identification approach. 
In any case, if the data-driven model is unable to produce a satisfactory prediction of the parameters $\boldsymbol{p^*}(t)$, its model structure should be revised.

% !TeX encoding = UTF-8
% !TeX spellcheck = en_US

\section{Case Studies}
\label{sec:caseStudies}

We present three case studies to demonstrate the effectiveness and versatility of the proposed incremental hybrid modeling approach.
\begin{itemize}
	\item \textbf{Case Study 1: Chemical Reactor} — This example illustrates the overall methodology in a controlled setting, highlighting the step-by-step application of the approach.
	\item \textbf{Case Study 2: Bioreactor} — This case focuses on a complex system with limited data availability, showcasing the robustness of the approach under realistic constraints.
	\item \textbf{Case Study 3: Research Plant with MPC} — {This real-world case demonstrates how the identified hybrid model can be integrated into an MPC framework for effective process control.}	
\end{itemize}
Before presenting the details of these case studies, we first describe the implementation strategies employed across the different scenarios.

\subsection{Implementation}
\label{sec:caseStudies:implementation}

We use two different implementations, due to the availability of the mechanistic model parts of the hybrid models.
In all cases, the machine learning model is identified with Python\footnote{Python, \url{www.python.org}, accessed: 14.04.2025} using Keras\footnote{Keras, \url{https://keras.io}, accessed: 14.04.2025} with the backend Tensorflow\footnote{TensorFlow, \url{www.tensorflow.org/}, accessed: 14.04.2025}.

Alternative 1:
In this approach, the hybrid model is implemented in gPROMS\footnote{gPROMS, \url{www.siemens.com/global/en/products/automation/industry-software/gproms-digital-process-design-and-operations.html}, accessed: 14.04.2025}. 
The dynamic optimization problem (Step 1) is solved using gPROMS, which employs a control vector parameterization approach \cite{Vassiliadis1994, Vassiliadis.1994b} to handle the dynamic optimization problem. 
The integration of the machine learning model into the hybrid model is achieved by importing the ONNX file from Keras into gPROMS using the adaptor:FO library in gPROMS.

Alternative 2:
In this approach, the hybrid model is implemented in Pyomo DAE\footnote{Pyomo DAE, \url{www.pyomo.org}, accessed: 14.04.2025} \cite{Nicholson2017}. 
The optimization problem is solved using full discretization \cite{Kameswaran2006, Biegler.2010} within Pyomo DAE. 
We utilize the widely-used nonlinear programming solver IPOPT version 3.11.1 to solve the dynamic parameter estimation problem \cite{Waechter2005}. 
The machine learning model is integrated into the hybrid model using the Python module OMLT\footnote{OMLT, \url{https://github.com/cog-imperial/OMLT}, accessed: 14.04.2025}.

\subsection{Illustrative Chemical Reactor Case Study}
\label{sec:caseStudies:chemicalReactor}

\begin{figure}[h!]
	\centering
	\begin{subfigure}{0.45\textwidth}
		\centering
		\includegraphics[width=\textwidth]{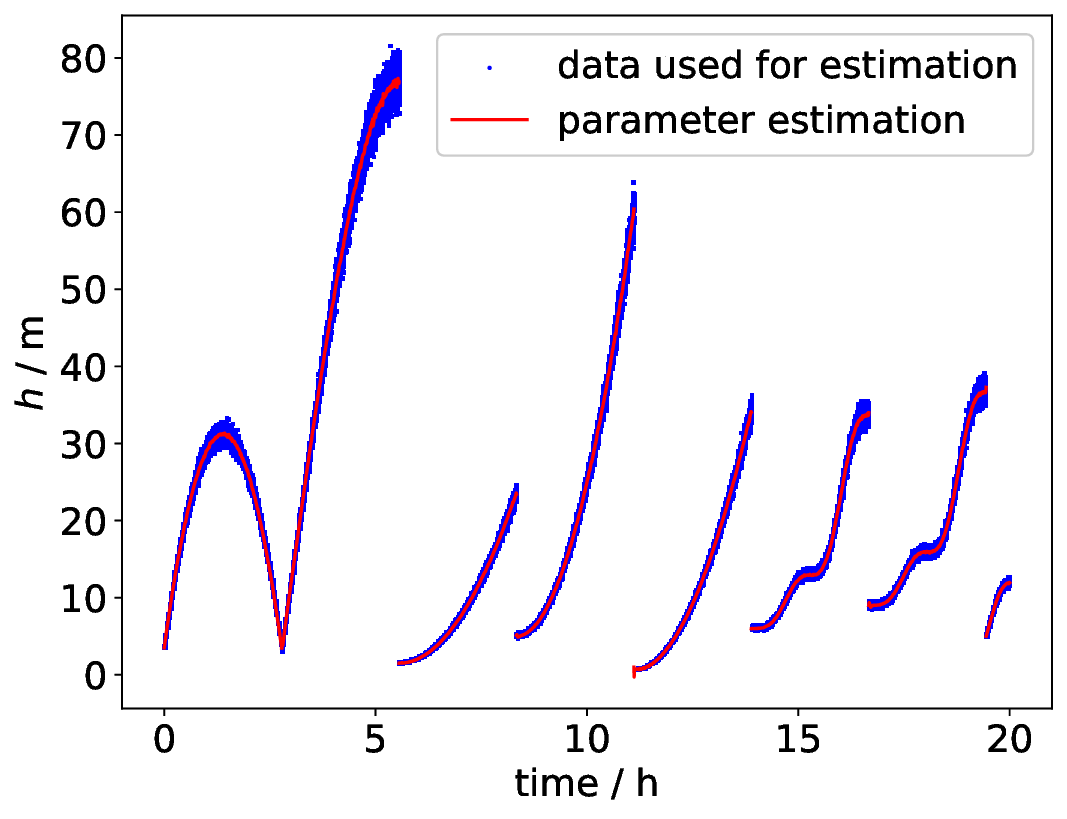}
		\caption{}
		\label{fig:chemicalReactor_comparison_height_parameter_est}
	\end{subfigure}
	\hfill
	\begin{subfigure}{0.45\textwidth}
		\centering
		\includegraphics[width=\textwidth]{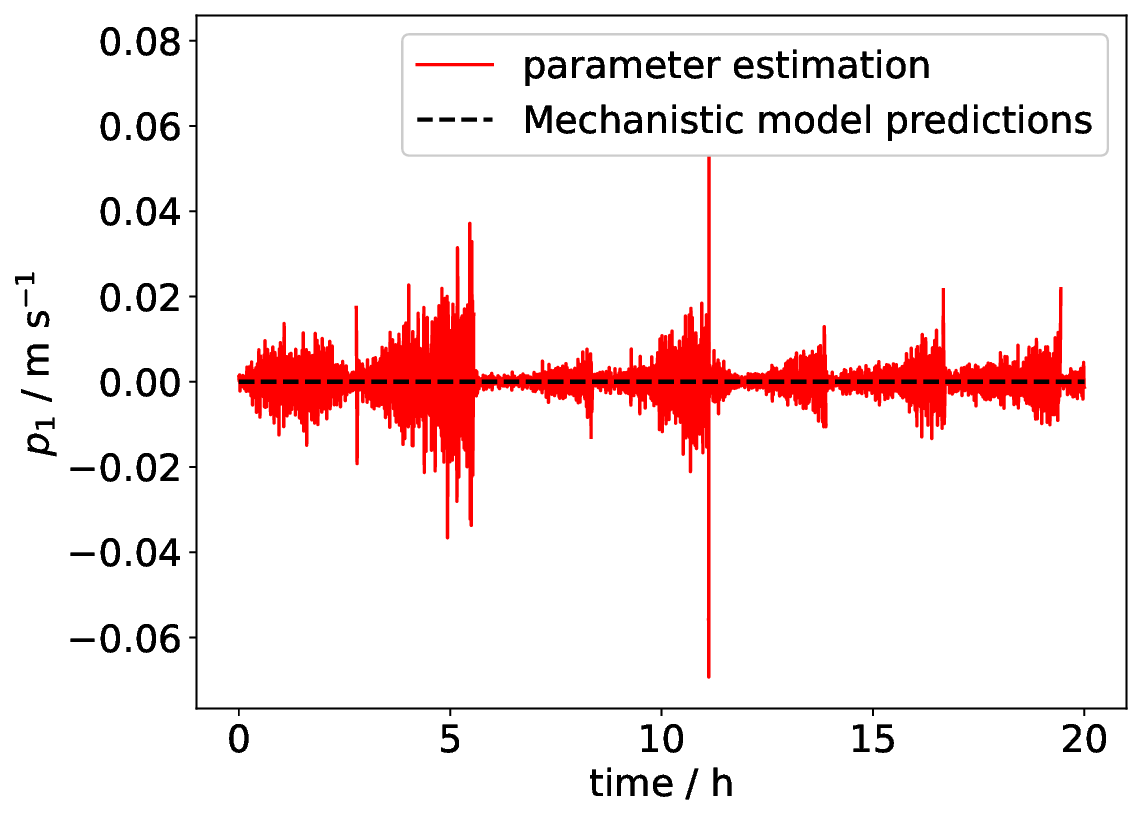}
		\caption{}
		\label{fig:chemicalReactor_comparison_k1_parameter_est}
	\end{subfigure}
	\begin{subfigure}{0.45\textwidth}
		\centering
		\includegraphics[width=\textwidth]{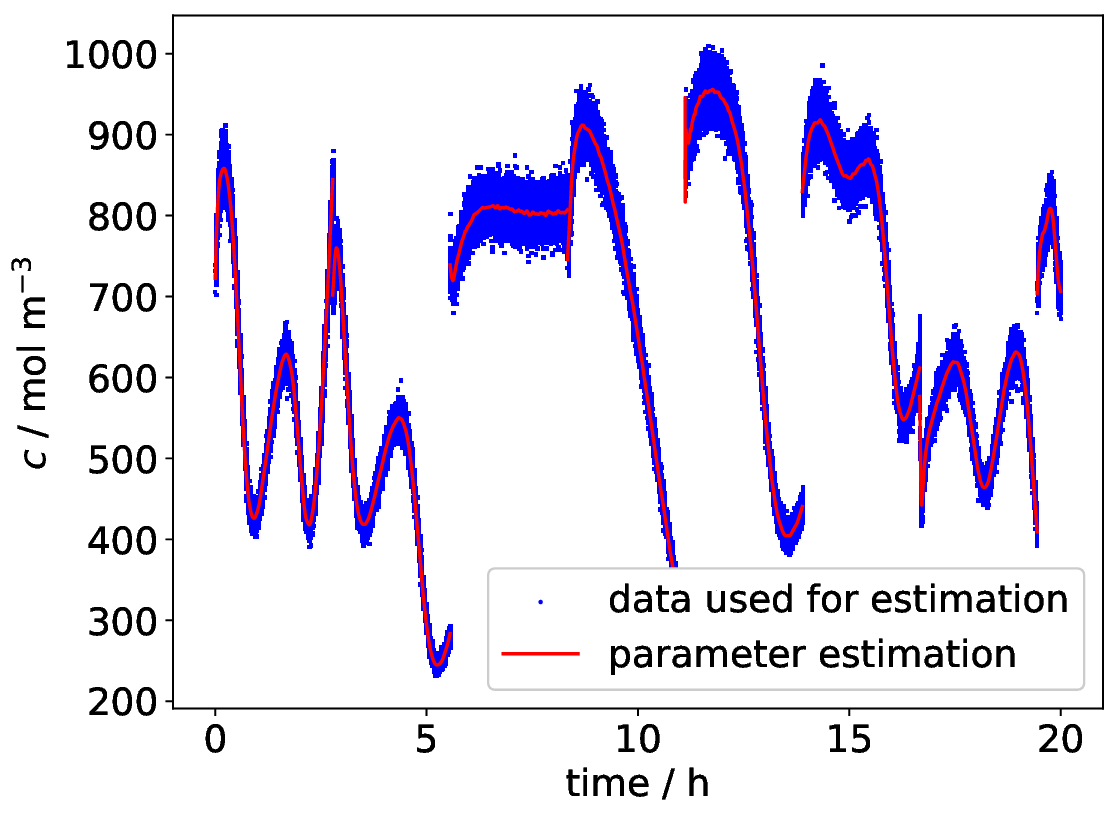}
		\caption{}
		\label{fig:chemicalReactor_comparison_concentration_parameter_est}
	\end{subfigure}
	\hfill
	\begin{subfigure}{0.45\textwidth}
		\centering
		\includegraphics[width=\textwidth]{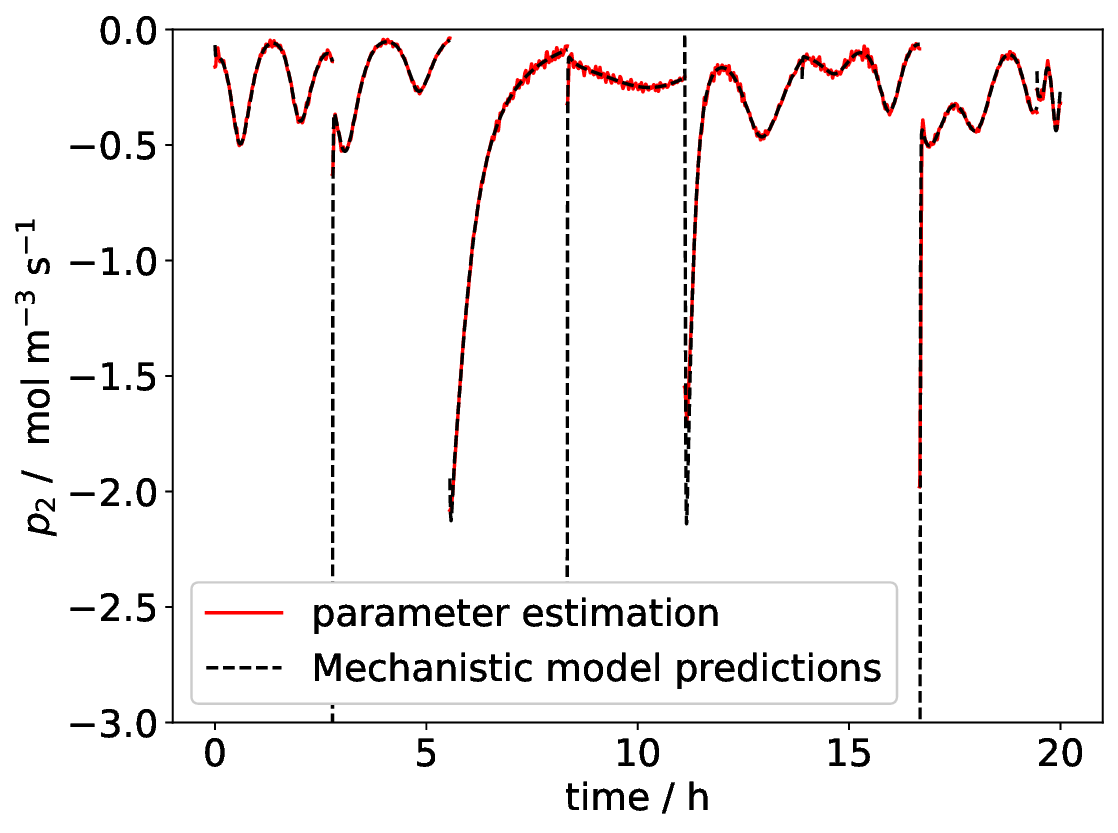}
		\caption{}
		\label{fig:chemicalReactor_comparison_k2_parameter_est}	
	\end{subfigure}
	\begin{subfigure}{0.45\textwidth}
		\centering
		\includegraphics[width=\textwidth]{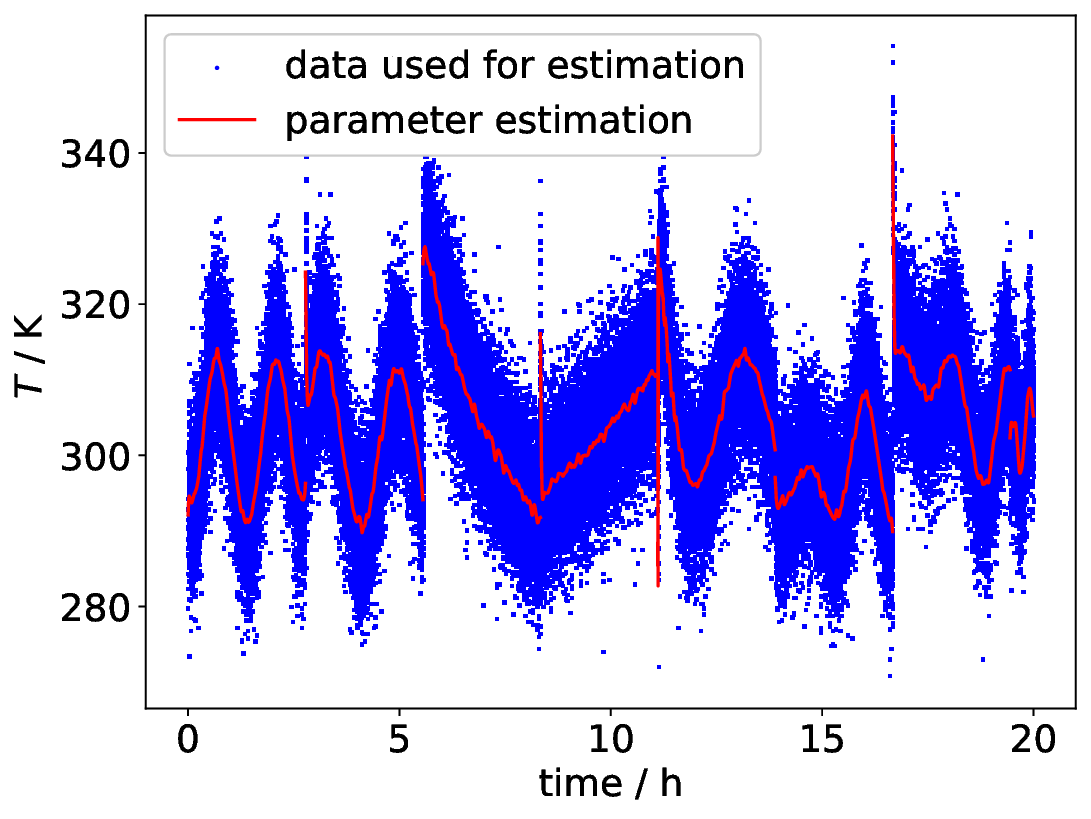}
		\caption{}
		\label{fig:chemicalReactor_comparison_temperature_parameter_est}
	\end{subfigure}
	\hfill
	\begin{subfigure}{0.45\textwidth}
		\centering
		\includegraphics[width=\textwidth]{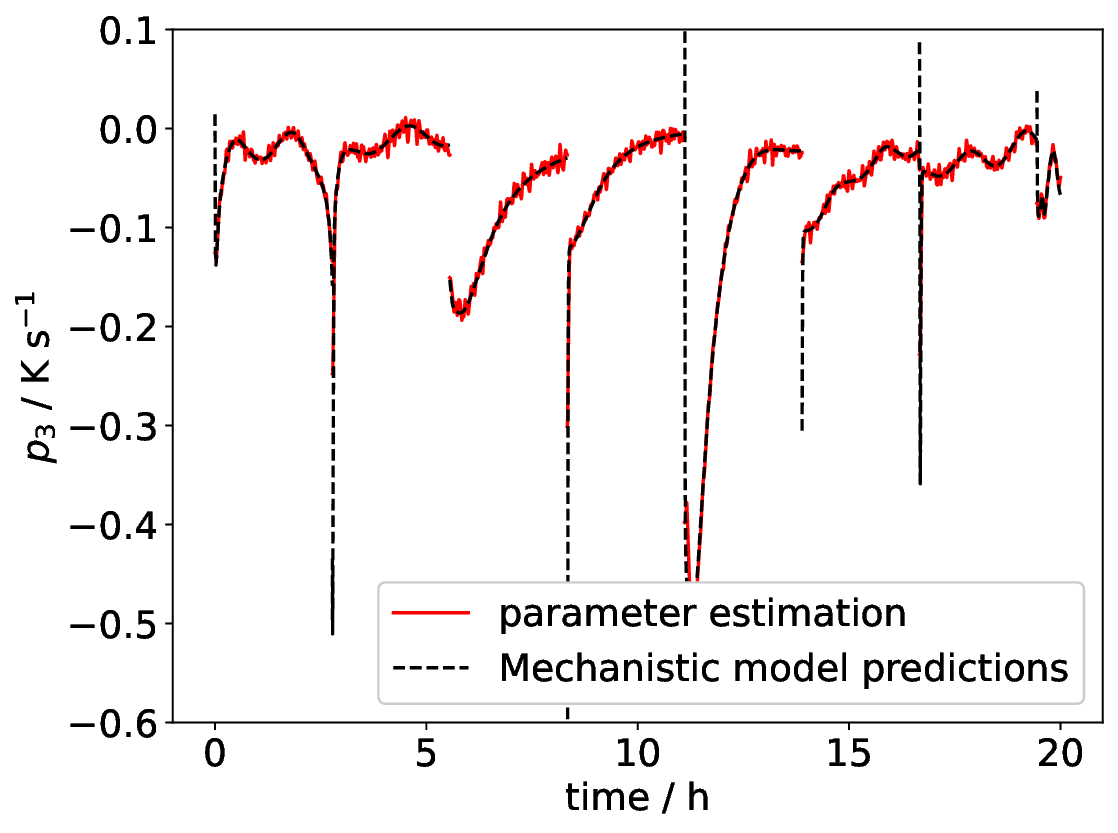}
		\caption{}
		\label{fig:chemicalReactor_comparison_k3_parameter_est}
	\end{subfigure}
	\caption{Results of dynamic parameter estimation problem (Step 1) for chemical reactor case study. Blue dots: Noisy pseudo-experimental measurement data. Red, solid lines: optimal profiles, i.e., solution of the parameter optimization problem. Black, dashed lines: {mechanistic model predictions} for parameters $p_1$, $p_2$, $p_3$. {For ease of presentation, each figure represents the concatenation of 8 separate experiments, each of 2.5 h duration but with different initial conditions and MV input profiles. (a) Height $h$. (b) Parameter $p_1$. (c) Concentration $c$. (d) Parameter $p_2$. (e) Temperature $T$. (f) Parameter $p_3$.}}
	\label{fig:chemicalReactor_comparison_parameter_est}
\end{figure}

This case study focuses on a continuous stirred tank reactor (CSTR).
A complete mechanistic model of this system as described in \cite{Rawlings2017} is defined by the following equations:
\begin{subequations}
	\begin{align}
		\frac{\mathrm{d}h}{\mathrm{d}t}(t) &= \frac{F_{0} - F_{\mathrm{out}}(t)}{\pi \cdot r^{2}} \label{eq:caseStudy1:mechanisticModel:a} \\
		\frac{\mathrm{d}c}{\mathrm{d}t}(t) &= \frac{F_{0} \cdot (c_{0} - c(t))}{\pi \cdot   r^{2} \cdot   h(t)} - k_{0} \cdot c(t) \cdot \mathrm{exp}\left(-\frac{E_{\mathrm{a}}}{R \cdot T(t)}\right) \label{eq:caseStudy1:mechanisticModel:b} \\
		\frac{\mathrm{d}T}{\mathrm{d}t}(t) &= \frac{F_{0} \cdot (T_{0} - T(t))}{\pi \cdot r^{2} \cdot h(t)} - \frac{\Delta H}{\rho \cdot C_{\mathrm{p}}} \cdot k_{0} \cdot c(t) \cdot \mathrm{exp}\left(-\frac{E_{a} }{R \cdot T(t)}\right) + \frac{2 \cdot U}{r \cdot \rho \cdot C_{\mathrm{p}}}  \left(T_{\mathrm{c}}(t) - T(t)\right), \label{eq:caseStudy1:mechanisticModel:c}
	\end{align}
	\label{eq:caseStudy1:mechanisticModel}
\end{subequations}
with the reactor height $h$, the feed flowrate $F_0$, the outlet flowrate $F_\mathrm{out}$, the concentration $c$, the inlet concentration $c_0$, the reactor radius $r$, the reaction constant $k_0$, the activation energy $E_\mathrm{a}$, the temperature $T$, the inlet temperature $T_0$, the reaction enthalpy $\Delta H$, the reaction constant $k_0$, the heat capacity $C_\mathrm{p}$, the density $\rho$, the heat transfer coefficient $U$, the reactor radius $r$, and the cooling water temperature $T_\mathrm{c}$.
The parameters of the CSTR model as presented in \cite{Rawlings2017} are shown in Table \ref{table:caseStudies:CSTR:parameters}.
The MVs of the CSTR are the outlet flowrate $F_\mathrm{out}$ and the cooling water temperature $T_\mathrm{c}$

\begin{table}[h!]
	\centering
	\caption{Parameters of the CSTR model as presented in \cite{Rawlings2017}.}
	\begin{tabular}{ l l l }
		\textbf{Parameter} & \textbf{Value} & \textbf{Unit} \\
		\hline
		$F_0$ & 0.1 & m$^3$/min \\
		$T_0$ & 350 & K \\
		$c_0$ & 1 & kmol/m$^3$ \\
		$r$ & 0.219 & m \\
		$k_0$ & $7.2 \times 10^{10}$ & min$^{-1}$ \\
		$E_\mathrm{R}$ & 8750 & K \\
		$U$ & 54.94 & kJ/min $\cdot$ m$^2 \cdot$ K \\
		$\rho$ & 1000 & kg/m$^3$ \\
		$C_p$ & 0.239 & kJ/kg$\cdot$K \\
		$\Delta H$ & $-5 \times 10^4$ & kJ/kmol \\ \hline
	\end{tabular}
	\label{table:caseStudies:CSTR:parameters}
\end{table}
To illustrate our incremental model identification methodology, we assume that only the first term on the right-hand side of each equation is known, whereas the heat transfer correlations and reaction kinetics are unknown or too difficult to model mechanistically. This leads to the following hybrid model structure:
\begin{subequations}
	\begin{align*}
		%   return m.d_h_d_t[t] == (m.F0[t] - model.F[t])/(np.pi*m.r**2)
		\frac{\mathrm{d}h}{\mathrm{d}t}(t) &= \frac{F_{0} - F_{\mathrm{out}}(t)}{\pi \cdot r^{2}} + p_{1}(t)\\
		% m.d_c_d_t[t] == m.F0[t]*(m.c0 - m.c[t])/(np.pi*m.r**2*m.h[t]) - m.k0*m.c[t]*exp(-m.EdR/(m.T[t]))
		\frac{\mathrm{d}c}{\mathrm{d}t}(t) &= \frac{F_{0} \cdot (c_{0} - c(t))}{\pi \cdot  r^{2} \cdot  h(t)} + p_{2}(t) \\
		% return m.d_T_d_t[t]== m.F0[t]*(m.T0 - m.T[t])/(np.pi*m.r**2*m.h[t]) - m.deltaH/(m.rho*m.Cp)*m.k0*m.c[t]*exp(-m.EdR/(m.T[t])) + 2*m.U/(m.r*m.rho*m.Cp)*(m.Tc[t]-m.T[t])
		\frac{\mathrm{d}T}{\mathrm{d}t}(t) &= \frac{F_{0} \cdot (T_{0} - T(t))}{\pi \cdot r^{2} \cdot h(t)} + p_{3}(t)
	\end{align*}
\end{subequations}

\subsubsection{Pseudo-Experimental Data Generation}
\label{sec:caseStudies:chemicalReactor:dataGeneration}

To emulate real-life data, the mechanistic model \eqref{eq:caseStudy1:mechanisticModel} is simulated with different initial conditions for the differential states and {with} different MV profiles (outlet flowrate $F_\mathrm{out}$ and cooling water temperature $T_\mathrm{c}$). In total, 8 sets of initial conditions and MV profiles were used, spanning {a total duration of} 20 hours. % with a {1-minute} discretization. 
{The MVs were varied in a piecewise constant manner over a uniform time grid with a 1-minute time interval. The values in each interval were taken to be random samples from normal distributions with means of 0.1 m$^3$/min for $F_{out}$ and 300 K for $T_c$, and standard deviations of 1 \% for both.}
{To create the pseudo-experimental measurements, we} add 2\% Gaussian white noise to the {predicted} values {of the differential states}.

\subsubsection{Step 1: Regularized Dynamic Parameter Estimation}
\label{sec:caseStudies:chemicalReactor:dataGeneration:step1}
In Step 1, we use implementation alternative 2 (cf. Section \ref{sec:caseStudies:implementation}) to solve the regularized dynamic parameter estimation problem \eqref{eq:parameter_estimation} and to obtain optimal profiles for the parameters $p_1$, $p_2$, and $p_3$. Figure \ref{fig:chemicalReactor_comparison_parameter_est} shows the optimal parameter profiles and corresponding output variable profiles. We see that the parameter estimation leads to a good fit for the output variables (differential states), i.e., the emulated measurement data, and that we are able to approximate the mechanistic (true) profiles for the parameters $p_1$, $p_2$, and $p_3$. In a real-life setting where the full mechanistic model is not available, the quality of the parameter estimation and the mechanistic model can be evaluated by analyzing how well the measurement data is approximated.

\subsubsection{Step 2: Correlation Analysis}

\begin{figure}[t]
	\centering
	\includegraphics[width=0.6\textwidth]{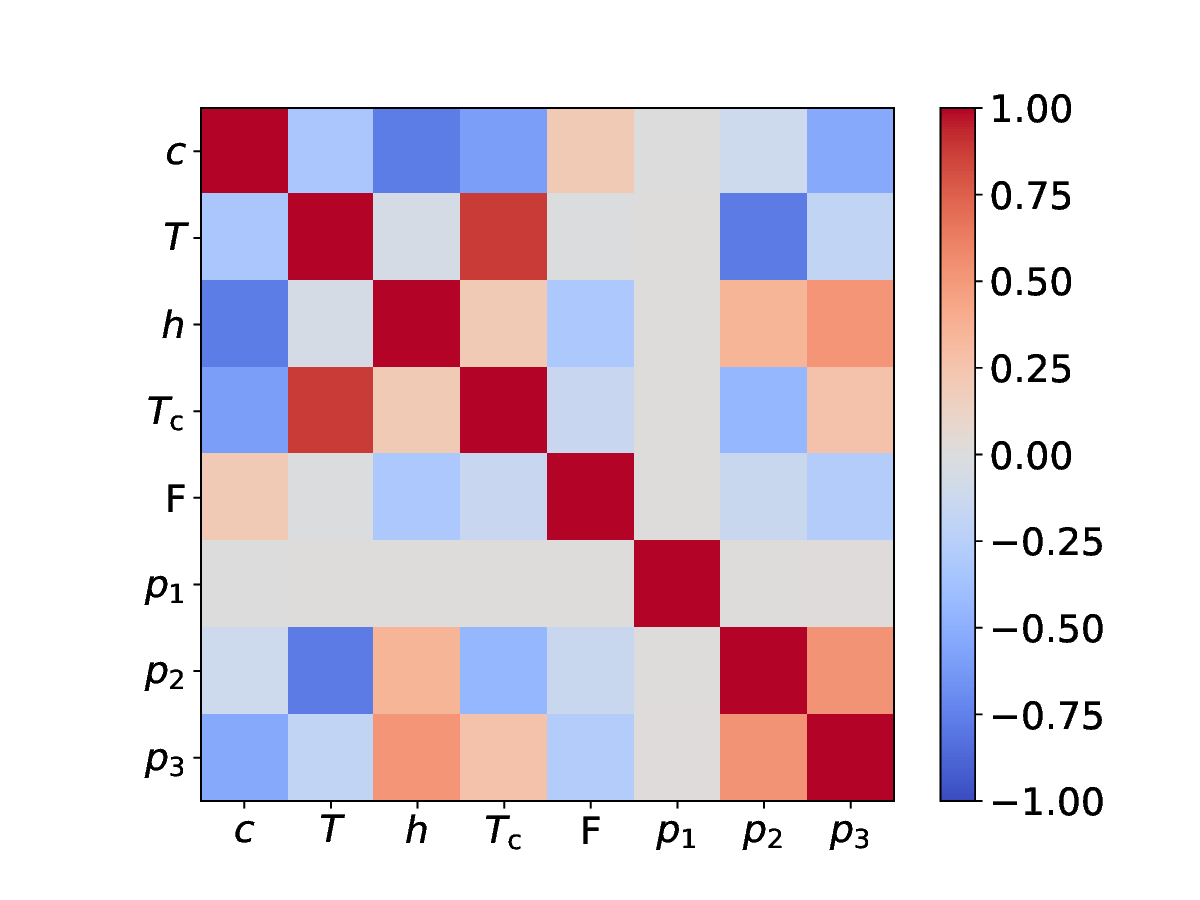}
	\caption{Pearson correlation matrix (Step 2) for chemical reactor case study.}
	\label{fig:correlation_matrix}
\end{figure}

In Step 2, we determine the correlations between the parameters $p_1$, $p_2$, and $p_3$, the differential states $h$, $c$, $T$, and the MVs $F_\mathrm{out}$, $T_\mathrm{c}$ of the hybrid model. Figure \ref{fig:correlation_matrix} shows the Pearson correlation matrix. We see that all differential states and inputs have high correlations with $p_2$ and $p_3$ and were thus used for the ANN training. Conversely, there is no high correlation of $p_1$ with the MVs and the differential states. Hence, no data-driven model will be used for $p_1$. 
Moreover, since the average value of the estimated $p_1$ is ${10}^{-6}$, $p_1$ is replaced by a constant value of zero.

\subsubsection{Step 3: Data-Driven Model Identification}

In Step 3, a data-driven model is identified to represent the parameters $p_2$ and $p_3$ as functions of all states and MVs. 
We train ANNs for both $p_2$ and $p_3$ with two hidden layers each comprising 4 neurons. 
The first hidden layer uses a tanh activation function while the second layer uses a linear activation. We do not show ANN training results here as the training in Keras/TensorFlow with standard settings was straightforward. 
Step 3 results in ANNs for the parameters $p_2$ and $p_3$, i.e., ${ML_1}(h(t), c(t), T(t), F_{\mathrm{out}}(t), T_{\mathrm{c}}(t))$ and ${ML_2}(h(t), c(t), T(t), F_{\mathrm{out}}(t), T_{\mathrm{c}}(t))$, respectively.

\subsubsection{Step 4: Hybrid Model Integration}

In Step 4, the parameters $p_2(t)$ and $p_3(t)$ are replaced by the trained data-driven models (ANNs) resulting from Step 3.
We receive, in turn, the following hybrid model:
\begin{subequations}
	\begin{align}
		\frac{\mathrm{d}h}{\mathrm{d}t}(t) &= \frac{F_{0} - F_{\mathrm{out}}(t)}{\pi \cdot r^{2}} \label{eq:caseStudy1:hybridModel:a} \\
		\frac{\mathrm{d}c}{\mathrm{d}t}(t) &= \frac{F_{0} \cdot (c_{0} - c(t))}{\pi \cdot  r^{2} \cdot  h(t)} + {ML_1}(h(t), c(t), T(t), F_{\mathrm{out}}(t), T_{\mathrm{c}}(t) ) \label{eq:caseStudy1:hybridModel:b} \\
		\frac{\mathrm{d}T}{\mathrm{d}t}(t) &= \frac{F_{0} \cdot (T_{0} - T(t))}{\pi \cdot r^{2} \cdot h(t)} + {ML_2}(h(t), c(t), T(t), F_{\mathrm{out}}(t), T_{\mathrm{c}}(t) ). \label{eq:caseStudy1:hybridModel:c}
	\end{align}
	\label{eq:caseStudy1:hybridModel}
\end{subequations}
To analyze the performance of the hybrid model, we simulate the hybrid model \eqref{eq:caseStudy1:hybridModel} for new MV profiles {different to those used for the generation of the pseudo-experimental data (cf. Section \ref{sec:caseStudies:chemicalReactor:dataGeneration}) but generated using the same randomized methodology} and compare the simulation results with those obtained using the original mechanistic model \eqref{eq:caseStudy1:mechanisticModel}, using the same initial states and MV profiles.

\begin{figure}[t]
	\centering
	\begin{subfigure}{0.45\textwidth}
		\centering
		\includegraphics[width=\textwidth]{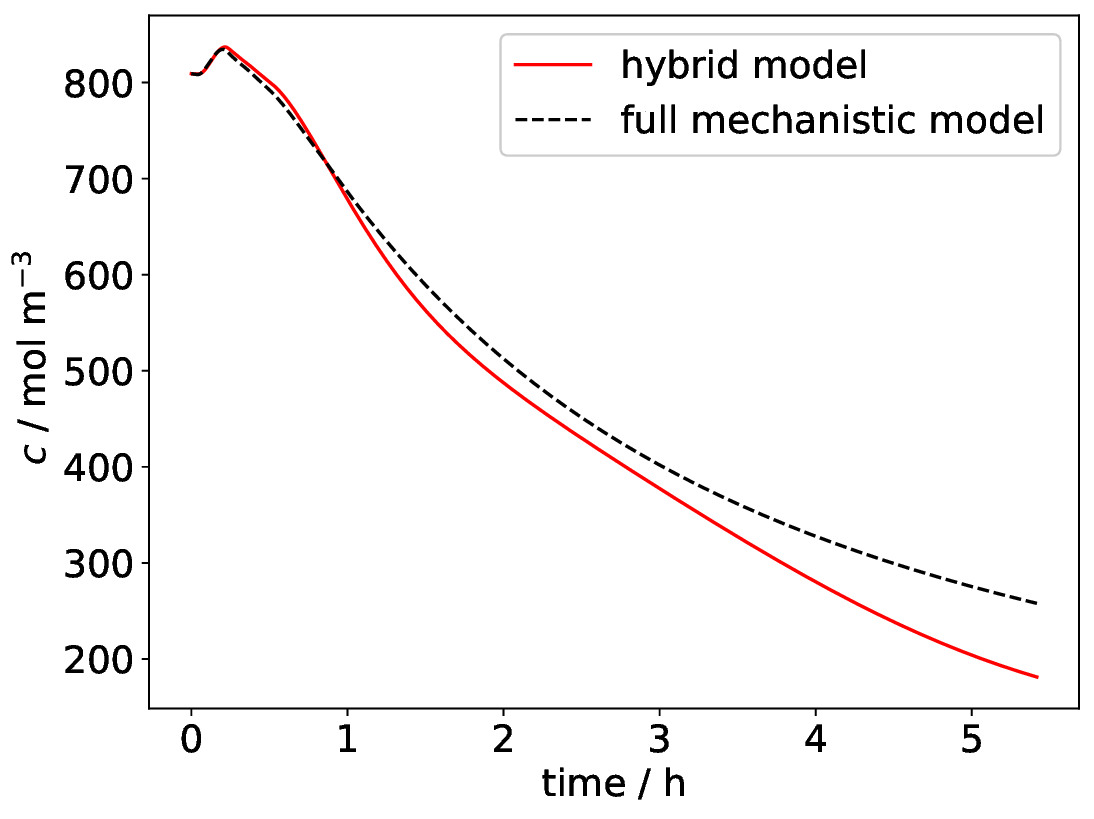}
		\caption{}
		\label{fig:chemicalReactor_comparison_concentration}
	\end{subfigure}
	\hfill
	\begin{subfigure}{0.45\textwidth}
		\centering
		\includegraphics[width=\textwidth]{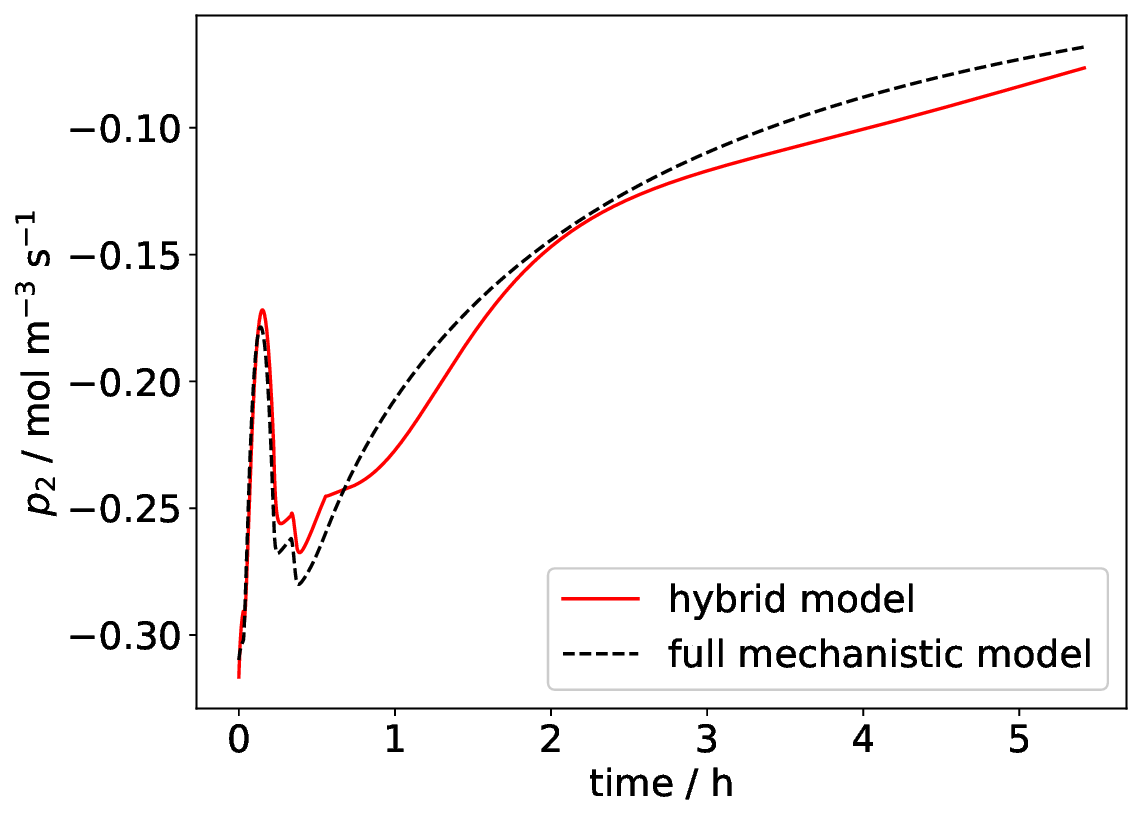}
		\caption{}
		\label{fig:chemicalReactor_comparison_k2}
	\end{subfigure}
	\begin{subfigure}{0.45\textwidth}
		\centering
		\includegraphics[width=\textwidth]{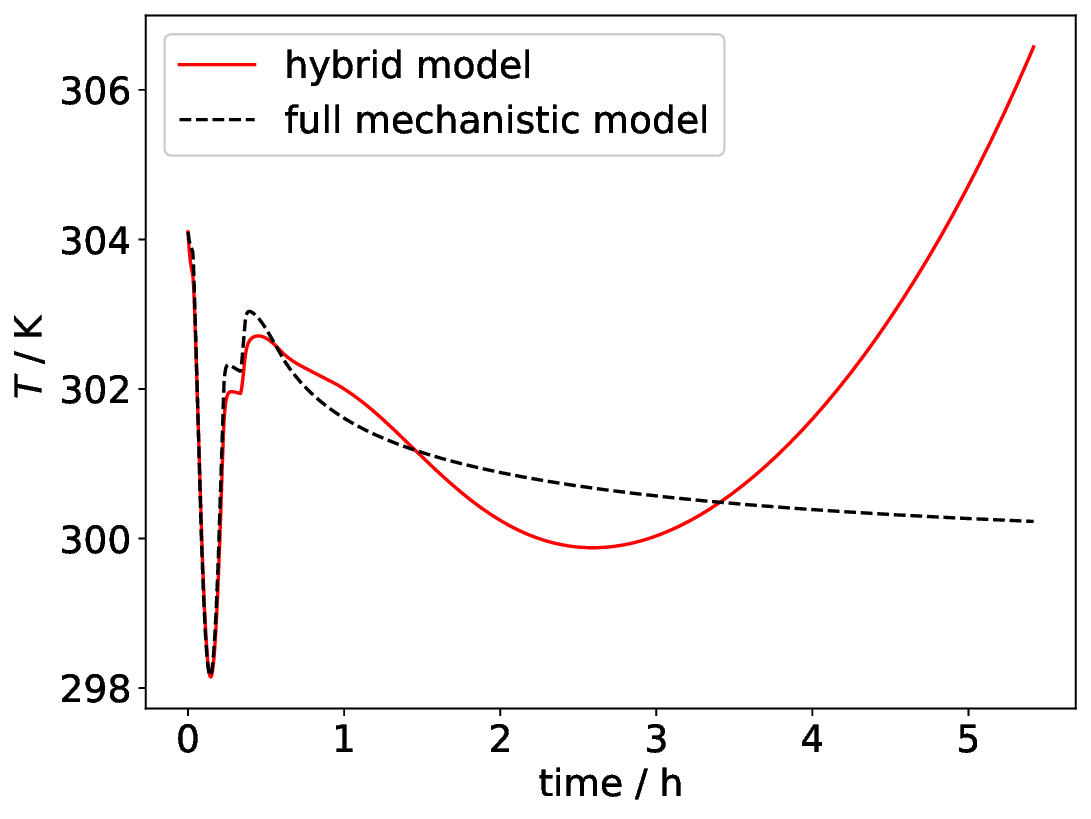}
		\caption{}
		\label{fig:chemicalReactor_comparison_temperature}
	\end{subfigure}
	\hfill
	\begin{subfigure}{0.45\textwidth}
		\centering
		\includegraphics[width=\textwidth]{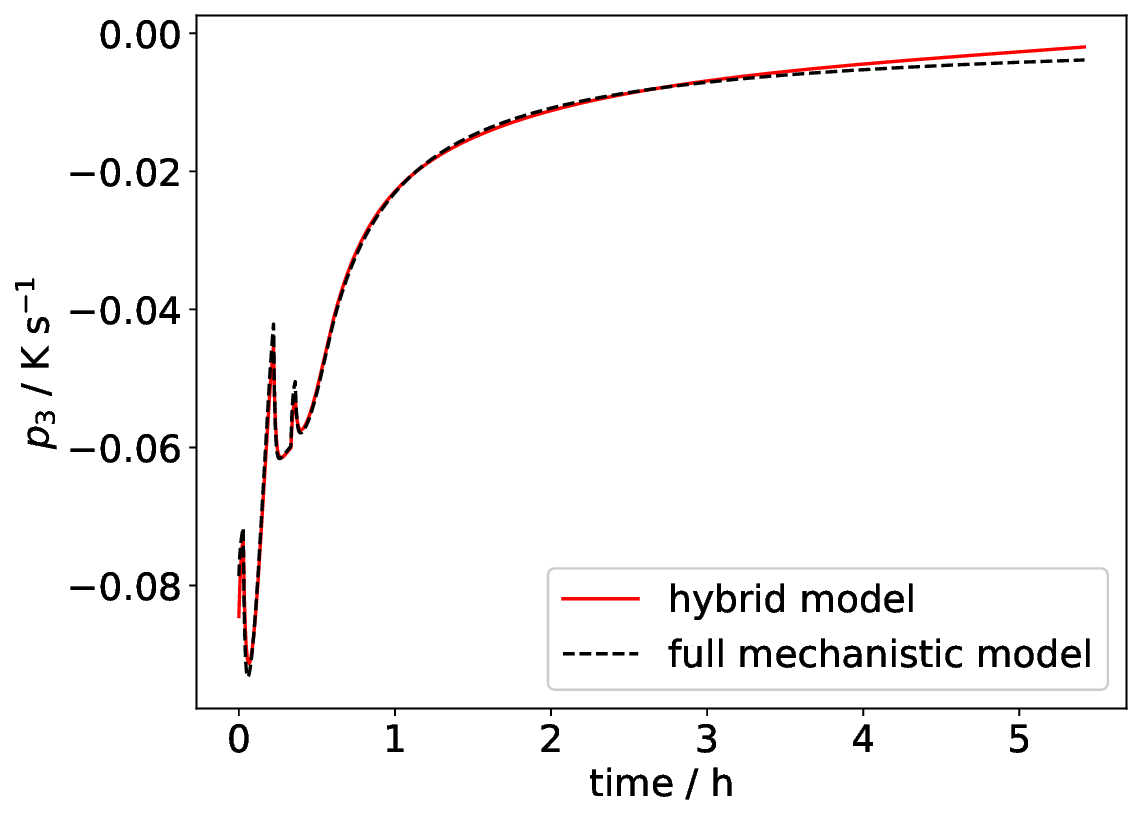}
		\caption{}
		\label{fig:chemicalReactor_comparison_k3}
	\end{subfigure}
	\caption{Simulation results with hybrid model compared to simulation results with mechanistic model for chemical reactor case study. Red, solid lines: simulation results with hybrid model. Black, dashed lines: simulation results with mechanistic model. (a) Concentration profiles. {(b) Parameter $p_2$ profiles. (c) Temperature profiles.} (d) Parameter $p_3$ profiles.}
	\label{fig:caseStudy1:evaluation}
\end{figure}

{
	Figure \ref{fig:caseStudy1:evaluation} shows a comparison of the predictions of the hybrid model against those of the mechanistic model. 
	As can be seen in Figures \ref{fig:chemicalReactor_comparison_k2} and \ref{fig:chemicalReactor_comparison_k3}, the time profiles of the parameters ${p}_2$ and ${p}_3$ computed by the data-driven element (neural network) within the hybrid model are reasonably close approximations to the true profiles of the corresponding terms in the mechanistic model. However, the corresponding profiles of the state profiles shown in Figures \ref{fig:chemicalReactor_comparison_concentration} and \ref{fig:chemicalReactor_comparison_temperature} exhibit larger divergences between hybrid and mechanistic model predictions. This is to be expected: a comparison of the forms of the two models (i.e. Equations \eqref{eq:caseStudy1:mechanisticModel:b} and \eqref{eq:caseStudy1:mechanisticModel:c} for the mechanistic model vs. Equations~\eqref{eq:caseStudy1:hybridModel:b} and \eqref{eq:caseStudy1:hybridModel:c} for the hybrid one) indicates that the state profile errors depend on the \textit{time integrals} of the errors in the corresponding parameters $\boldsymbol{p}$. 
	Consequently, over long simulation times, errors in the hybrid model’s predictions of the states may accumulate significantly even if the parameters $\boldsymbol{p}$ are predicted well. 
	Moreover, whether or not the magnitudes of these state errors are increasing or decreasing at a particular point in time will depend on whether the state and parameter errors have the same sign (positive or negative).
	This can clearly be seen from a comparison of Figures~\ref{fig:chemicalReactor_comparison_temperature} and \ref{fig:chemicalReactor_comparison_k3}. 
	Overall, in applications where the hybrid model is deployed online over extended periods of time, the use of state estimation techniques is essential for correcting the state trajectories by taking account of real-time measurements.
}

\subsubsection{{Influence of Regularization in Dynamic Parameter Estimation Problem (Step 1)}}
\label{sec:caseStudies:chemicalReactor:regularization}

\begin{figure}[h!t]
	\centering
	\begin{subfigure}{0.45\textwidth}
		\centering
		\includegraphics[width=\textwidth]{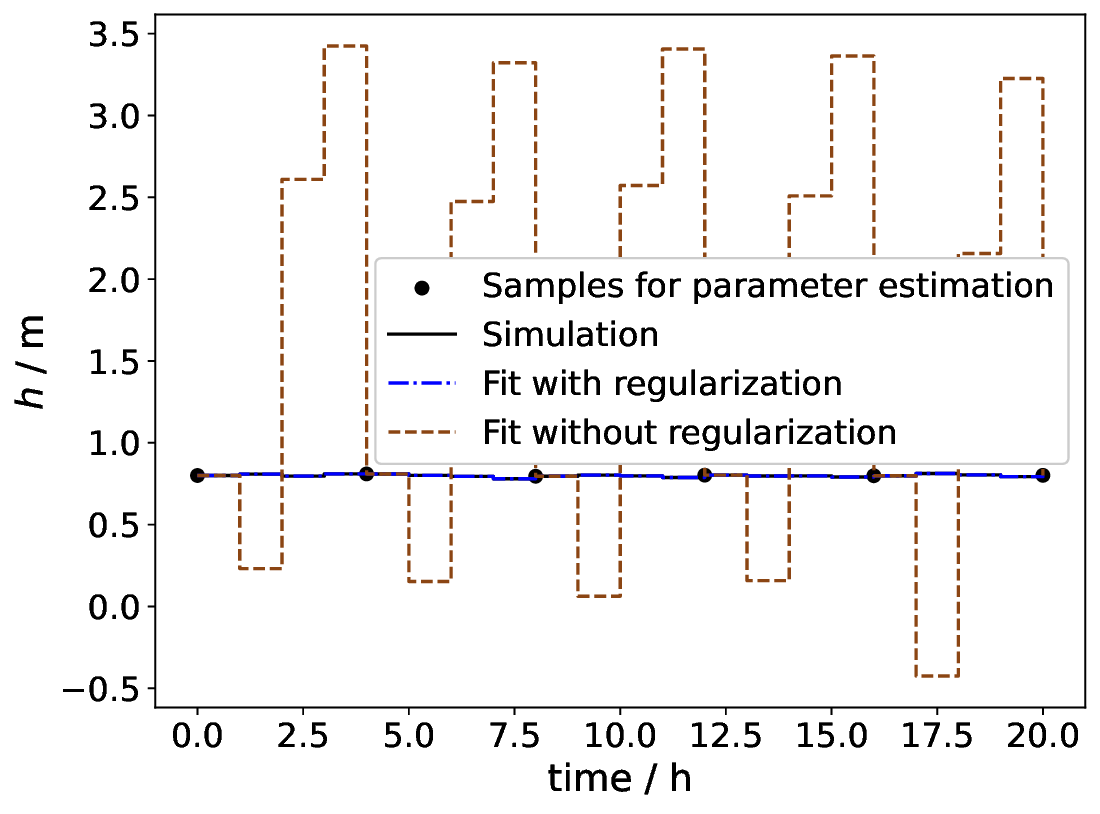}
		\caption{}
		\label{fig:caseStudy1:regularizationTest:figs:fit_h}
	\end{subfigure}
	\hfill
	\begin{subfigure}{0.45\textwidth}
		\centering
		\includegraphics[width=\textwidth]{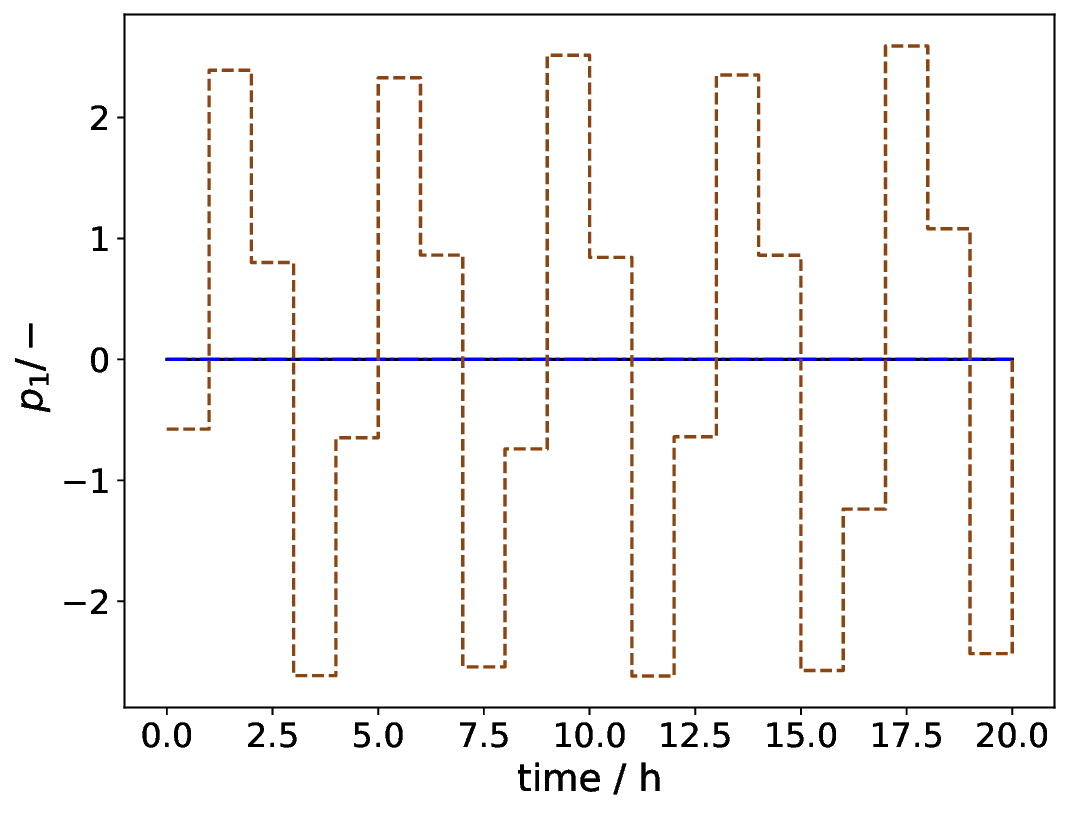}
		\caption{}
		\label{fig:caseStudy1:regularizationTest:figs:fit_k_1}
	\end{subfigure}
	\begin{subfigure}{0.45\textwidth}
		\centering
		\includegraphics[width=\textwidth]{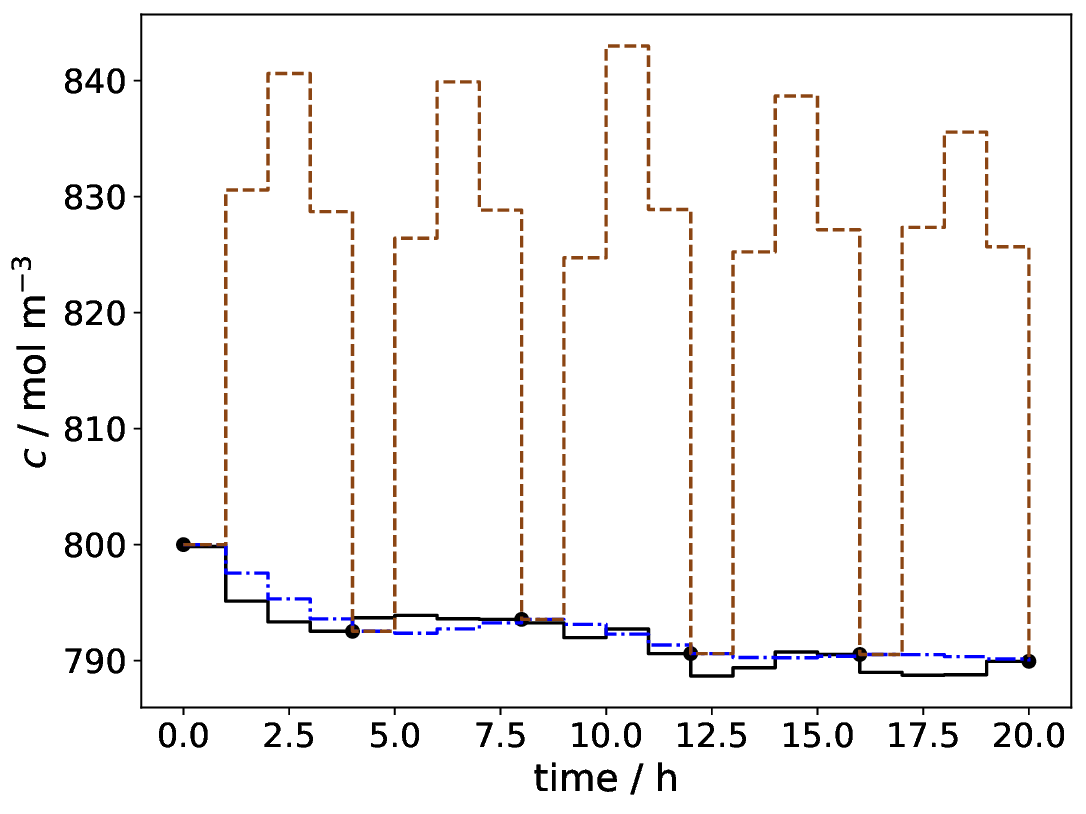}
		\caption{}
		\label{fig:caseStudy1:regularizationTest:figs:fit_c}
	\end{subfigure}
	\hfill
	\begin{subfigure}{0.45\textwidth}
		\centering
		\includegraphics[width=\textwidth]{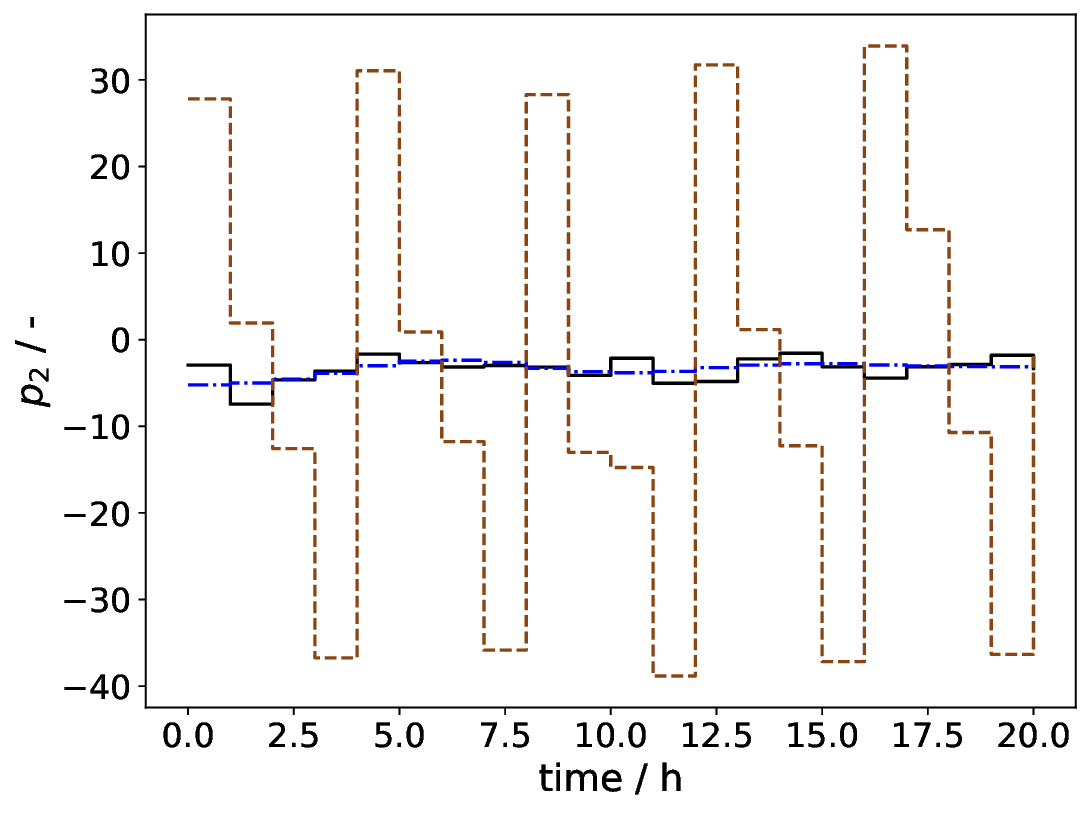}
		\caption{}
		\label{fig:caseStudy1:regularizationTest:figs:fit_k_2}
	\end{subfigure}
	\begin{subfigure}{0.45\textwidth}
		\centering
		\includegraphics[width=\textwidth]{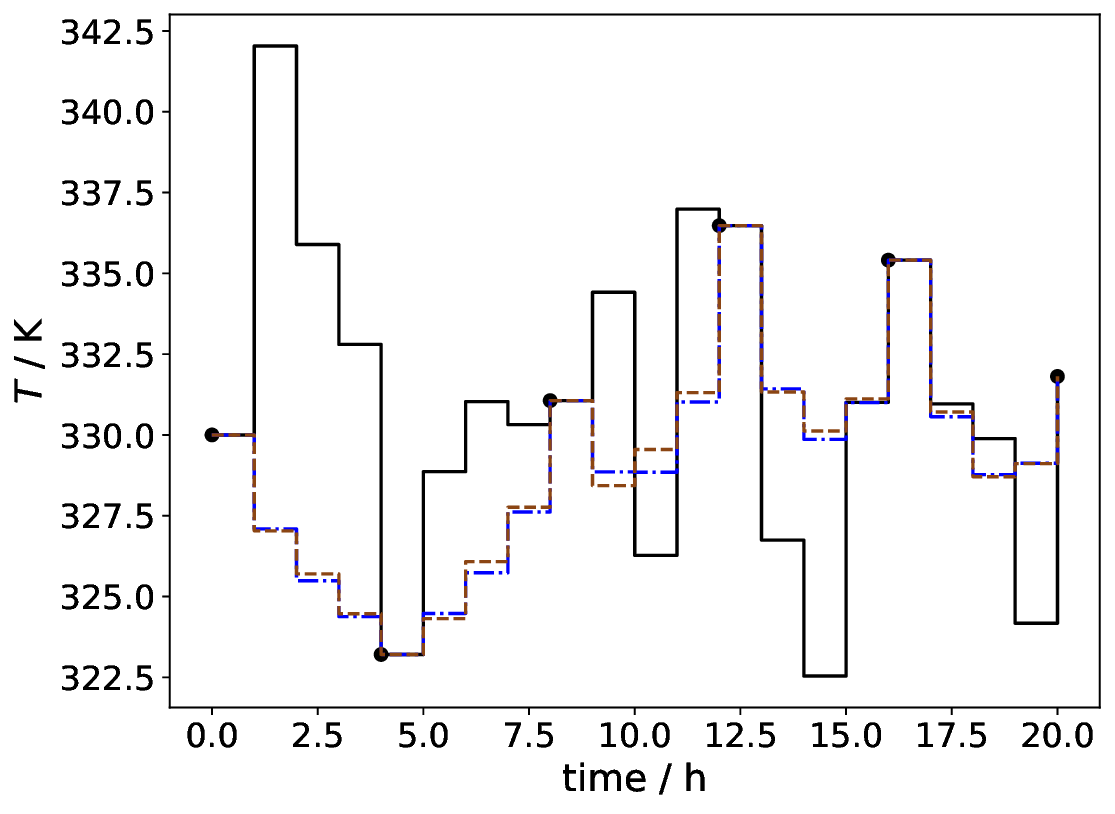}
		\caption{}
		\label{fig:caseStudy1:regularizationTest:figs:fit_T}
	\end{subfigure}
	\hfill
	\begin{subfigure}{0.45\textwidth}
		\centering
		\includegraphics[width=\textwidth]{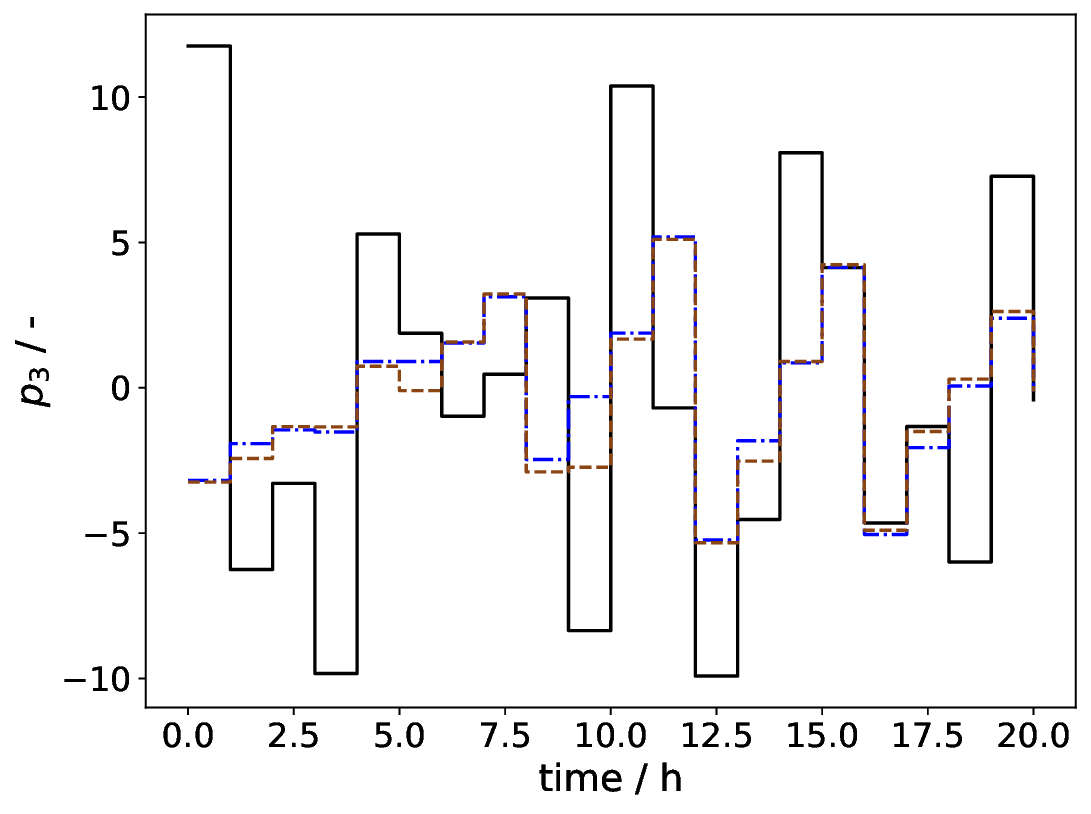}
		\caption{}
		\label{fig:caseStudy1:regularizationTest:figs:fit_k_3}
	\end{subfigure}
	\caption{{Solution of dynamic parameter estimation (step 1) with the regularization term (blue, dash-dotted lines) and without the regularization term (brown, dashed lines) in the objective function \eqref{eq:parameter_estimation:obj}. The measurement samples used for parameter estimation are illustrated by black dots. (a) Height profile. The correct profiles (determined via simulation of the original model) are illustrated by black, solid lines. (b) Parameter $p_1$ profile. (c) Concentration profile. (d) Parameter $p_2$ profile. (e) Temperature profile. (f) Parameter $p_3$ profile.}}
	\label{fig:caseStudy1:regularizationTest:figs}
\end{figure}

{
	To illustrate the impact of the regularization term $R(\boldsymbol{p}(t))$ on the results the dynamic parameter estimation problem 	\eqref{eq:parameter_estimation}, we consider a slightly modified version of the problem solved in Section \ref{sec:caseStudies:chemicalReactor:dataGeneration:step1}. 
	In particular, to emulate the issues arising from sparsity of experimental data, the estimation problem takes account of only a small subset of the measurement points. 
	Moreover, in the interests of clarity of results presentation, the uniform grid used for the piecewise constant representation of the parameter functions $p(t)$ involves time intervals of duration 1 h instead of 1 minute as is the case in Section \ref{sec:caseStudies:chemicalReactor:dataGeneration:step1}.
}

{
	The results with and without the regularization term (cf. Equation \eqref{eq:regularization}) are shown in Figure~\ref{fig:caseStudy1:regularizationTest:figs}. 
	It can be seen that, while both approaches yield a good fit to the sparse measurement data $\boldsymbol{\tilde{z}}_j$ (cf. Figures~\ref{fig:caseStudy1:regularizationTest:figs:fit_k_1}, \ref{fig:caseStudy1:regularizationTest:figs:fit_k_2} and \ref{fig:caseStudy1:regularizationTest:figs:fit_k_3}), significant differences emerge in the smoothness and physical plausibility of the resulting profiles $\boldsymbol{p}(t)$ and $\boldsymbol{x}(t)$. Without regularization, the estimated parameter trajectories exhibit strong eratic fluctuations and deviate notably from the true simulation profiles. 
	In contrast, the inclusion of the regularization term $R(\boldsymbol{p}(t))$ (cf. Equation~\eqref{eq:regularization}) leads to smoother parameter and state trajectories that more closely follow the underlying simulation data. This highlights the benefit of regularization in promoting physically meaningful solutions, especially when experimental data is limited.
}

{It is worth noting that, for the examples considered in this paper, we assumed a diagonal structure for the matrices $\boldsymbol{W}_k^R$ used in the regularization term of Equation 	\eqref{eq:regularization}, and adopted an empirical approach for their determination. In particular, starting with matrices with zero or very small elements in the diagonal, we solved the regularized dynamic parameter estimation problem \eqref{eq:parameter_estimation} and observed the resulting time profiles for $\boldsymbol{p}(t)$. If one or more of the resulting $\boldsymbol{p}$ was found to exhibit erratic behavior, we increased the corresponding elements in $\boldsymbol{W}_k^R$ and repeated the solution.}

\subsubsection{Conclusion}

To summarize this case study, the incremental modeling approach allows for the development of an accurate hybrid model.  
The heat transfer correlations and reaction kinetics can be modeled by the data-driven model parts without the need for expert knowledge or structural knowledge of the mechanistic equations.

\subsection{Bioreactor Case Study}
\label{sec:caseStudies:bioreactor}

\begin{figure}[t]
	\centering
	\begin{subfigure}{0.45\textwidth}
		\centering
		\includegraphics[width=\textwidth]{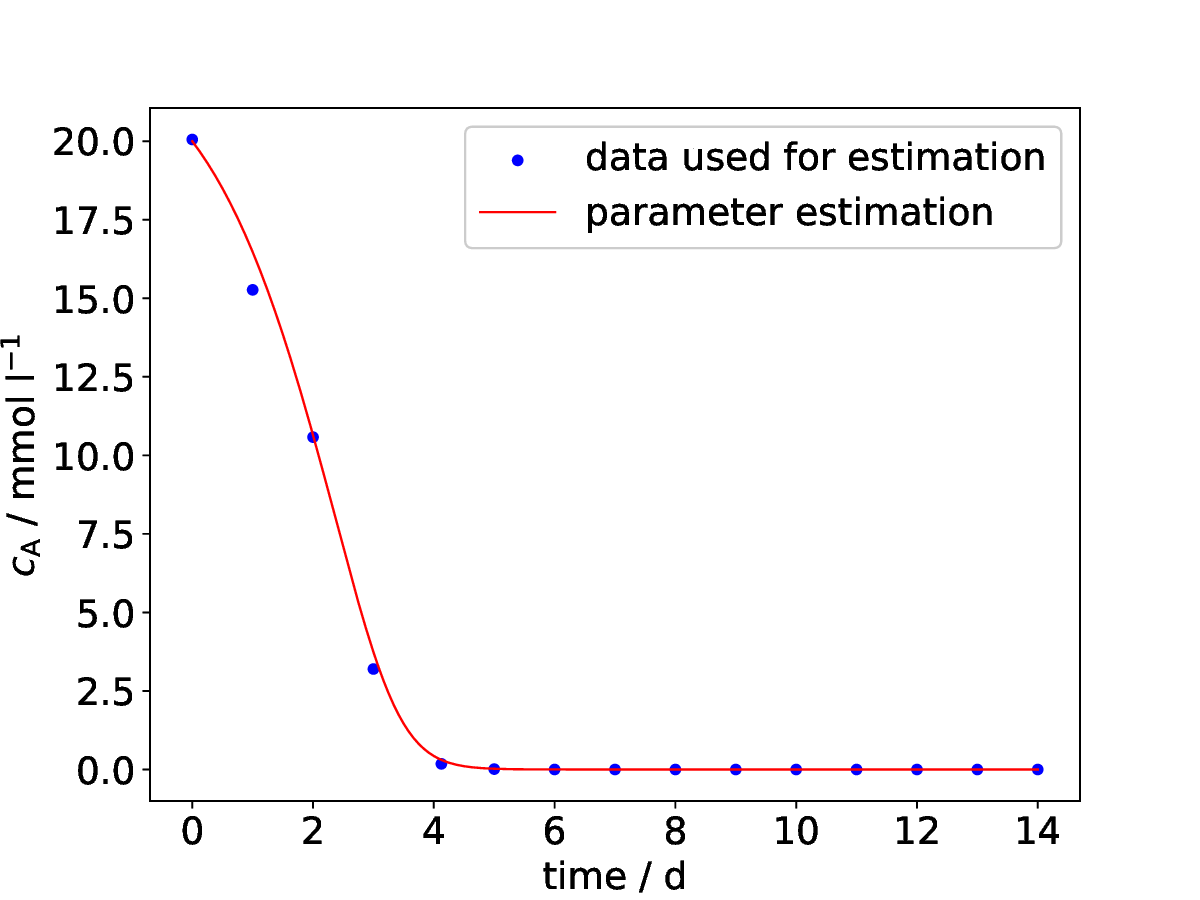}
		\caption{}
		\label{fig:bioReactor_aspara20}
	\end{subfigure}
	\hfill
	\begin{subfigure}{0.45\textwidth}
		\centering
		\includegraphics[width=\textwidth]{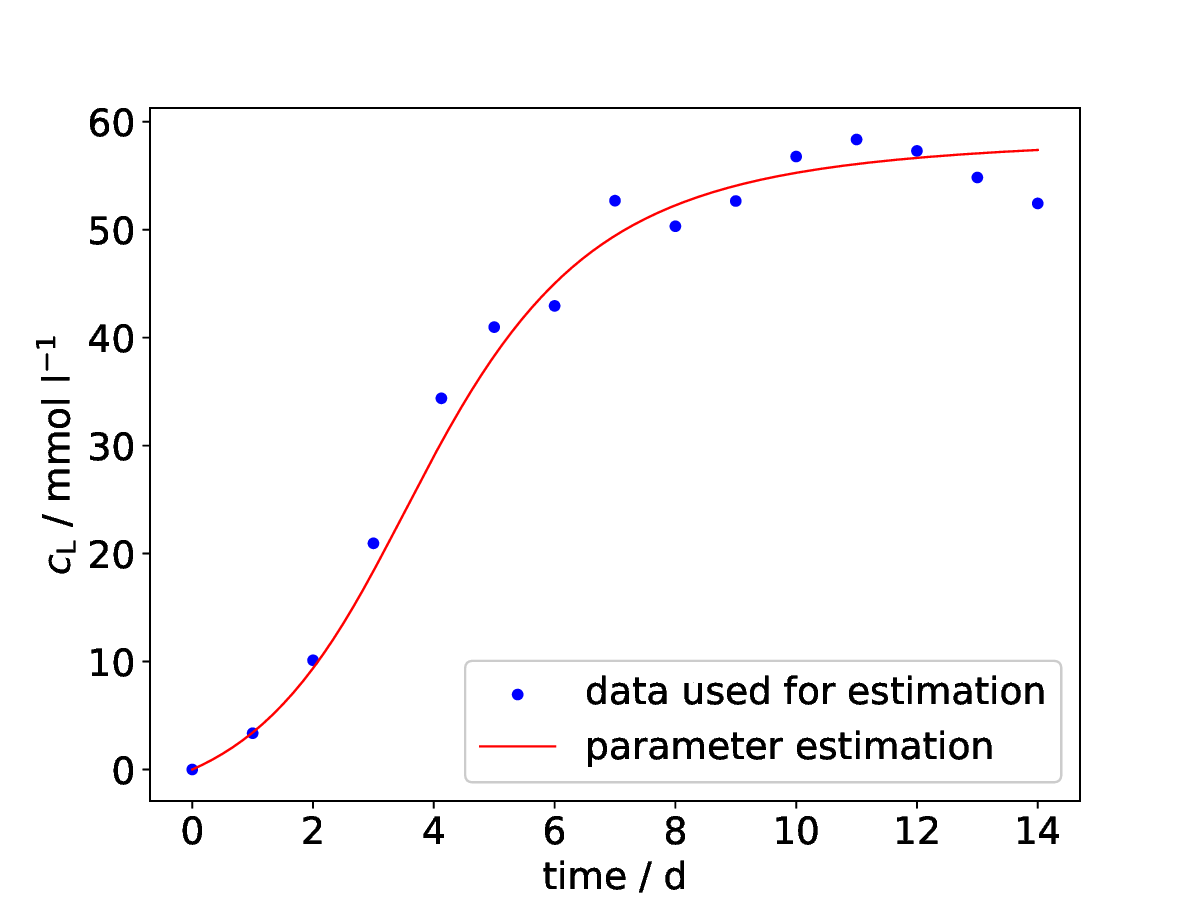}
		\caption{}
		\label{fig:bioReactor_lactate20}
	\end{subfigure}
	\begin{subfigure}{0.45\textwidth}
		\centering
		\includegraphics[width=\textwidth]{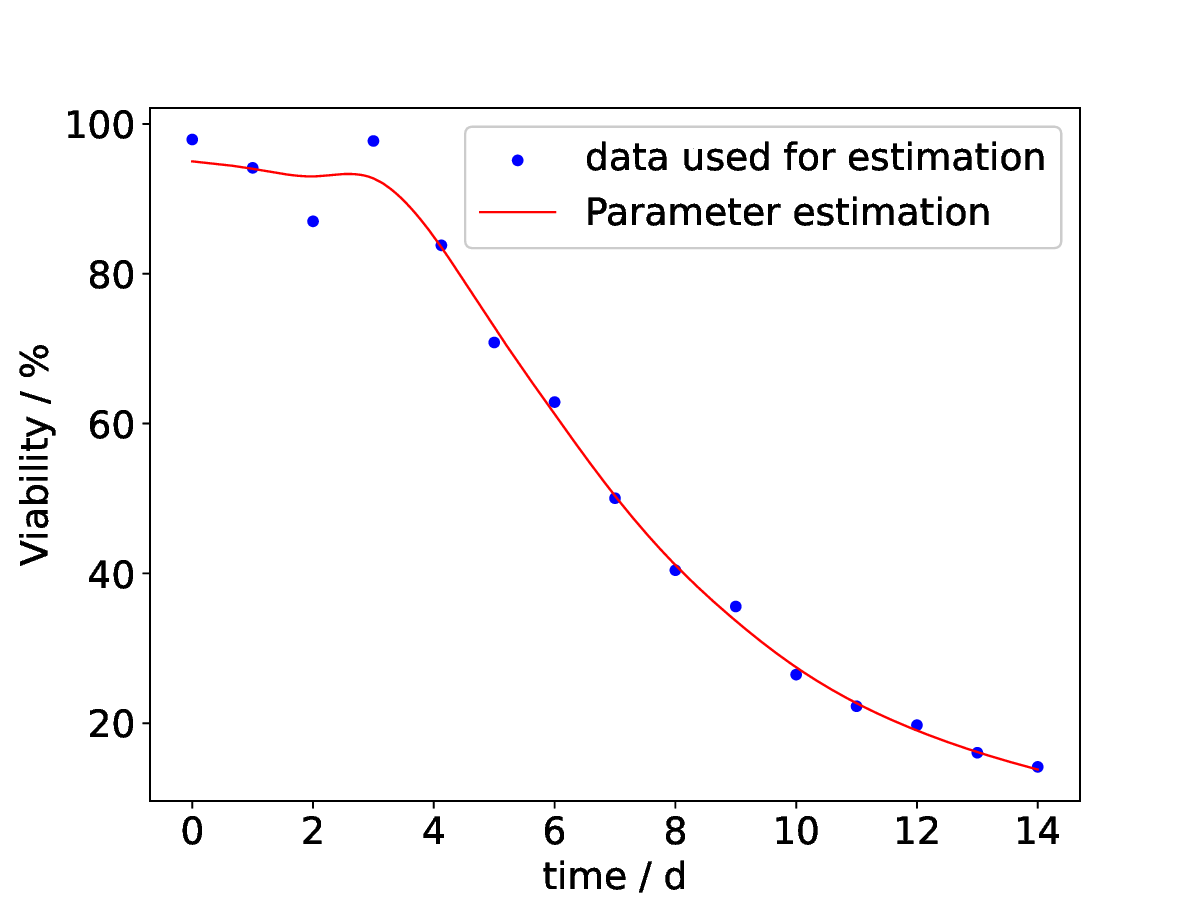}
		\caption{}
		\label{fig:bioReactor_cell20}
	\end{subfigure}
	\hfill
	\begin{subfigure}{0.45\textwidth}
		\centering
		\includegraphics[width=\textwidth]{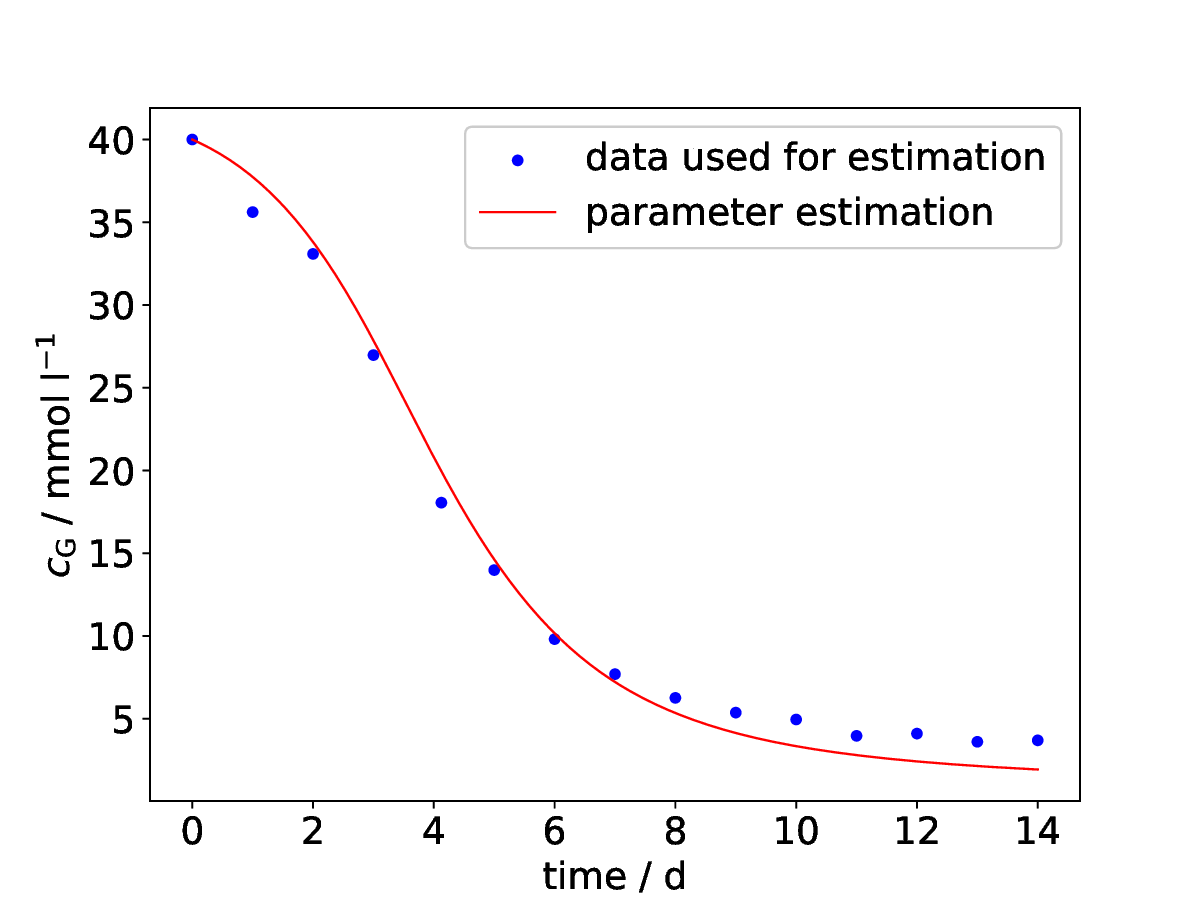}
		\caption{}
		\label{fig:bioReactor_glucose20}
	\end{subfigure}
	\caption{Results of dynamic parameter estimation problem (Step 1) for bioreactor case study. Blue points: noisy measurement data. Red, solid lines: optimal profiles resulting from the solution of the parameter estimation problem. (a) Asparagine concentration. (b) Lactate concentration. (c) Cell viability. (d) Glucose concentration.}
	\label{fig:bioReactor_comparison_diff_states_aspara20}
\end{figure}

We consider a complex bioreactor system in this case study. 
The mechanistic model of the bioreactor is {implemented} in gPROMS FormulatedProducts\footnote{gPROMS FormulatedProducts, version 2023.2.0, \url{www.siemens.com/global/en/products/automation/industry-software/gproms-digital-process-design-and-operations/gproms-modelling-environments/gproms-formulatedproducts.html}, accessed: 26.05.2025} and consists {of 30 differential and over 2500 algebraic equations}. 
We use implementation alternative 1 for this case study (cf. Section \ref{sec:caseStudies:implementation}).

{
	As is often the case in modeling of bioreactors, the main gap in mechanistic knowledge is related to the characterization of the underlying biological phenomena. 
	Here, guided by domain-specific knowledge, we assume that the major such uncertainty is associated with the cell growth and cell death mechanisms. 
	We therefore treat the specific cell growth and death rates as the parameters $p_1$ and $p_2$ that need to be characterized via data-driven elements in our hybrid model.	
}

\subsubsection{Pseudo-Experimental Data Generation}

To generate pseudo-experimental data, we simulate the mechanistic bioreactor model using three distinct initial conditions for the asparagine concentration \( c_\mathrm{A0} \): 1, 7, and 20 mmol/L. The output variables include glucose concentration \( c_\mathrm{G} \), lactate concentration \( c_\mathrm{L} \), asparagine concentration \( c_\mathrm{A} \), and cell viability \( v \), sampled daily over a 14-day period. To emulate realistic measurement noise, we add 5\% Gaussian white noise to the output values.

\subsubsection{Step 1: Regularized Dynamic Parameter Estimation}

In the first step, we formulate and solve the regularized dynamic parameter estimation problem in gPROMS. Figure \ref{fig:bioReactor_comparison_diff_states_aspara20} illustrates the results, showing that despite the limited amount of data, we obtain smooth profiles for the output variables (differential states) with a good fit.

\subsubsection{Step 2: Correlation Analysis}

\begin{figure}[t]
	\centering
	\includegraphics[width=0.6\textwidth]{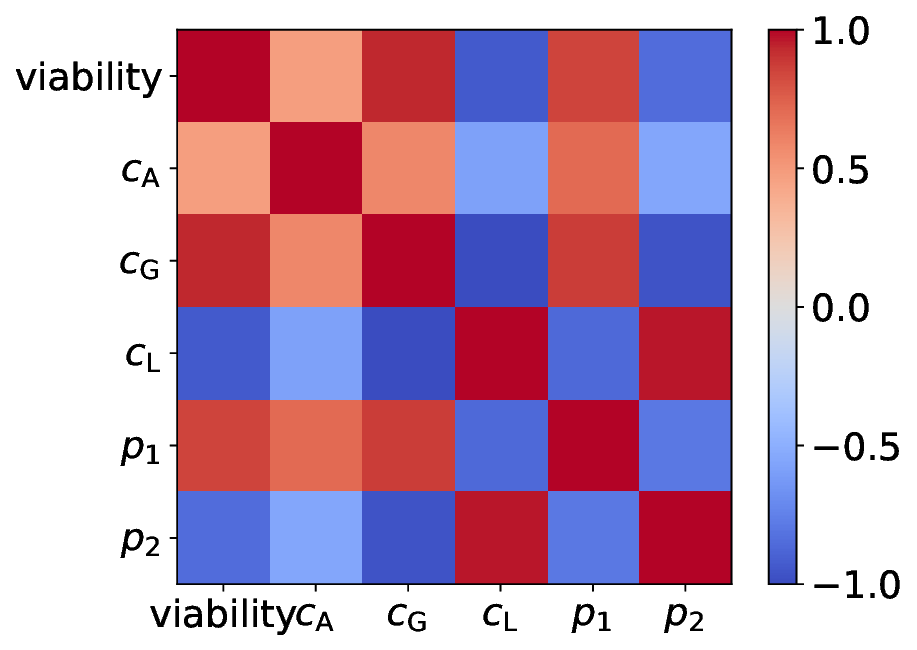}
	\caption{Pearson correlation matrix (Step 2) for bioreactor case study.}
	\label{fig:correlation_matrix_training_data_bio}
\end{figure}

In Step 2, we calculate the Pearson correlation matrix to analyze the correlations between the parameters $p_1$ and $p_2$, the differential states, and the manipulated variables (MVs). 
Figure \ref{fig:correlation_matrix_training_data_bio} shows the correlation matrix, indicating that the considered differential states have high correlations with the parameters $p_1$ and $p_2$. 
Therefore, in Step 3, those differential states are considered as inputs for the data-driven models for $p_1$ and $p_2$.

\subsubsection{Step 3: Data-Driven Model Identification}

In Step 3, we identify data-driven models (artificial neural networks, ANNs) for the parameters $p_1$ and $p_2$ as functions of the differential states. The ANNs have two hidden layers with a leaky ReLU activation function and a width of 10 nodes. Additionally, a dropout layer with a dropout rate of 0.1 is added. The training is straightforward with a standard setup in Keras/TensorFlow. As a result, we obtain the data-driven models $ML_1(c_\mathrm{G}, c_\mathrm{L}, c_\mathrm{A}, v)$ and $ML_2(c_\mathrm{G}, c_\mathrm{L}, c_\mathrm{A}, v)$ to replace the parameters $p_1$ and $p_2$, respectively.

\subsubsection{Step 4: Hybrid Model Integration}

\begin{figure}[t]
	\centering
	\begin{subfigure}{0.45\textwidth}
		\centering
		\includegraphics[width=\textwidth]{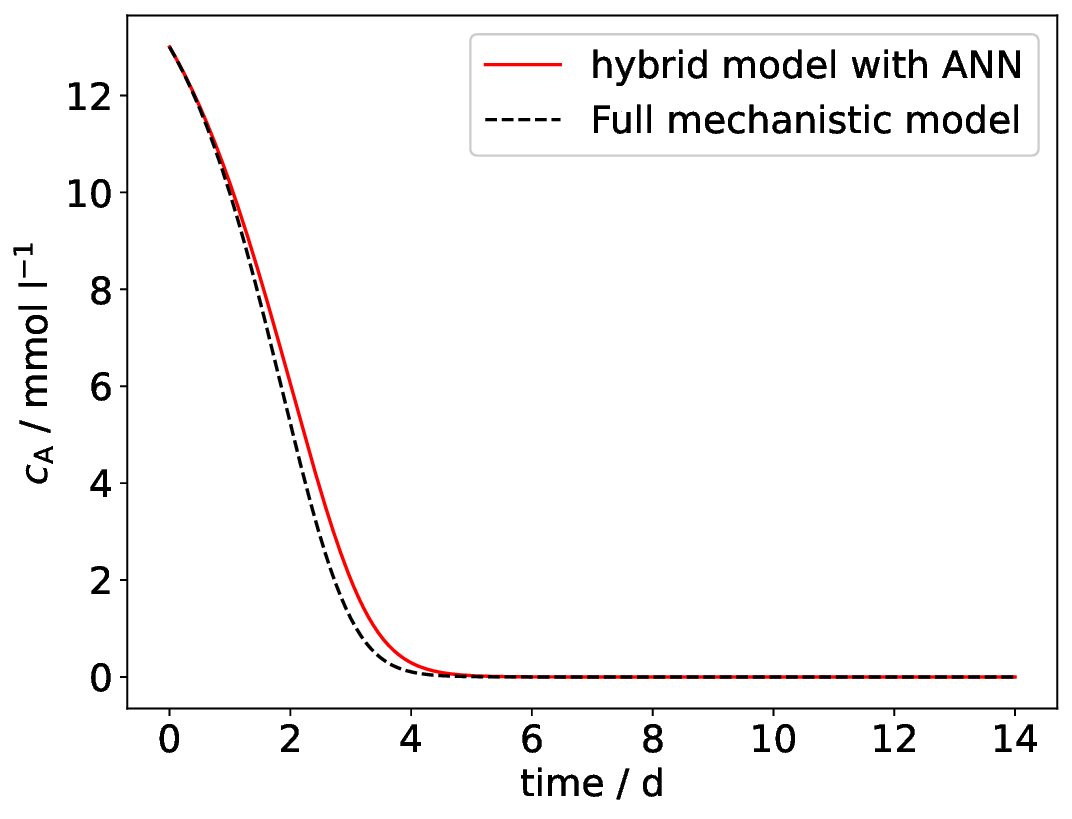}
		\caption{}
		\label{fig:bioReactor_aspara13}
	\end{subfigure}
	\hfill
	\begin{subfigure}{0.45\textwidth}
		\centering
		\includegraphics[width=\textwidth]{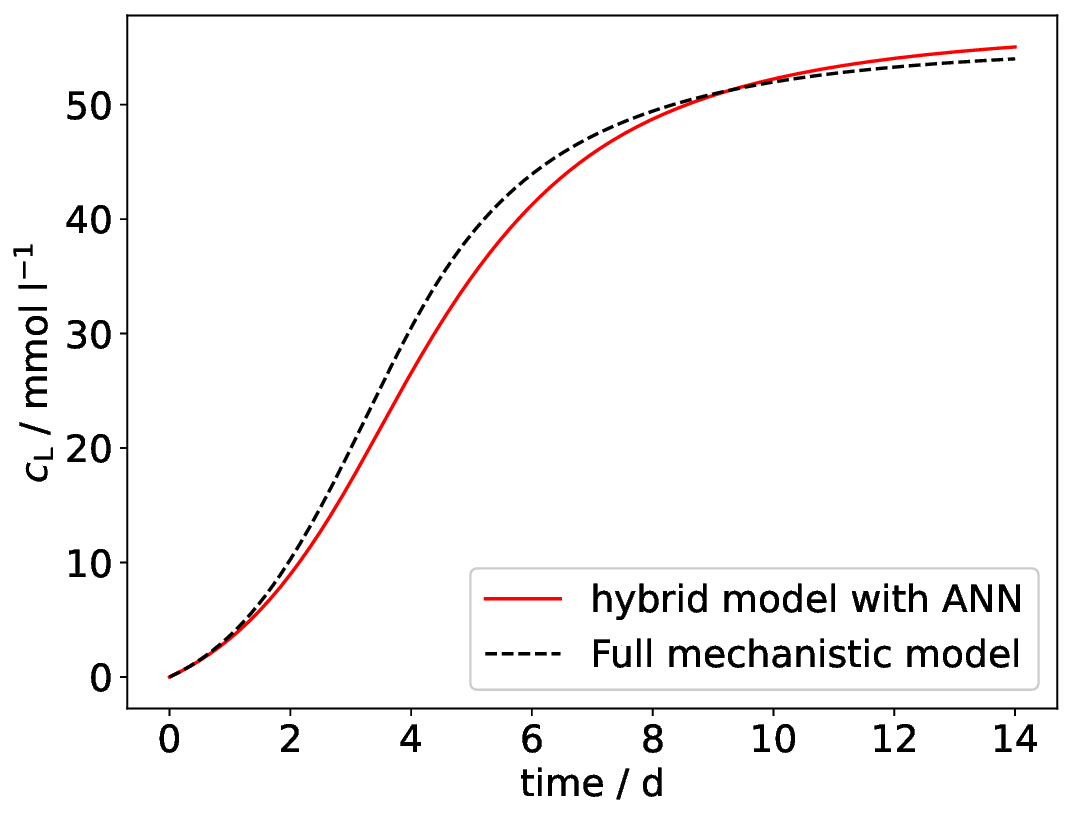}
		\caption{}
		\label{fig:bioReactor_lactate13}
	\end{subfigure}
	\begin{subfigure}{0.45\textwidth}
		\centering
		\includegraphics[width=\textwidth]{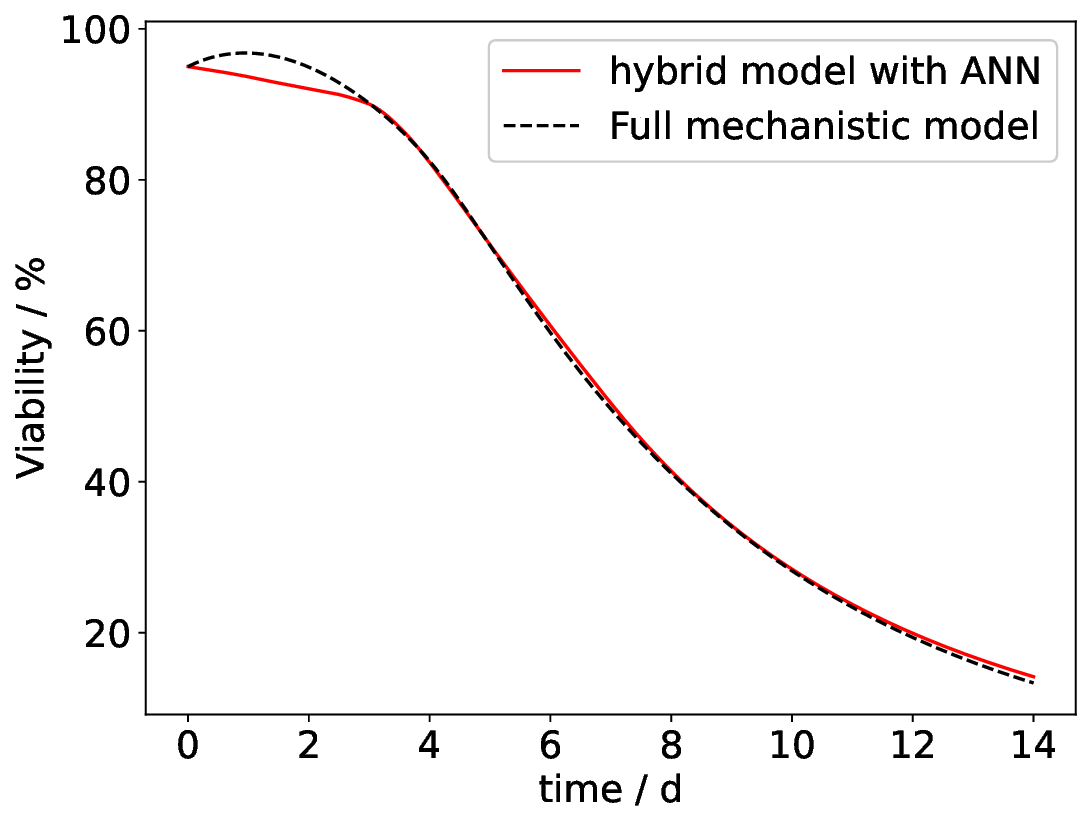}
		\caption{}
		\label{fig:bioReactor_cell13}
	\end{subfigure}
	\hfill
	\begin{subfigure}{0.45\textwidth}
		\centering
		\includegraphics[width=\textwidth]{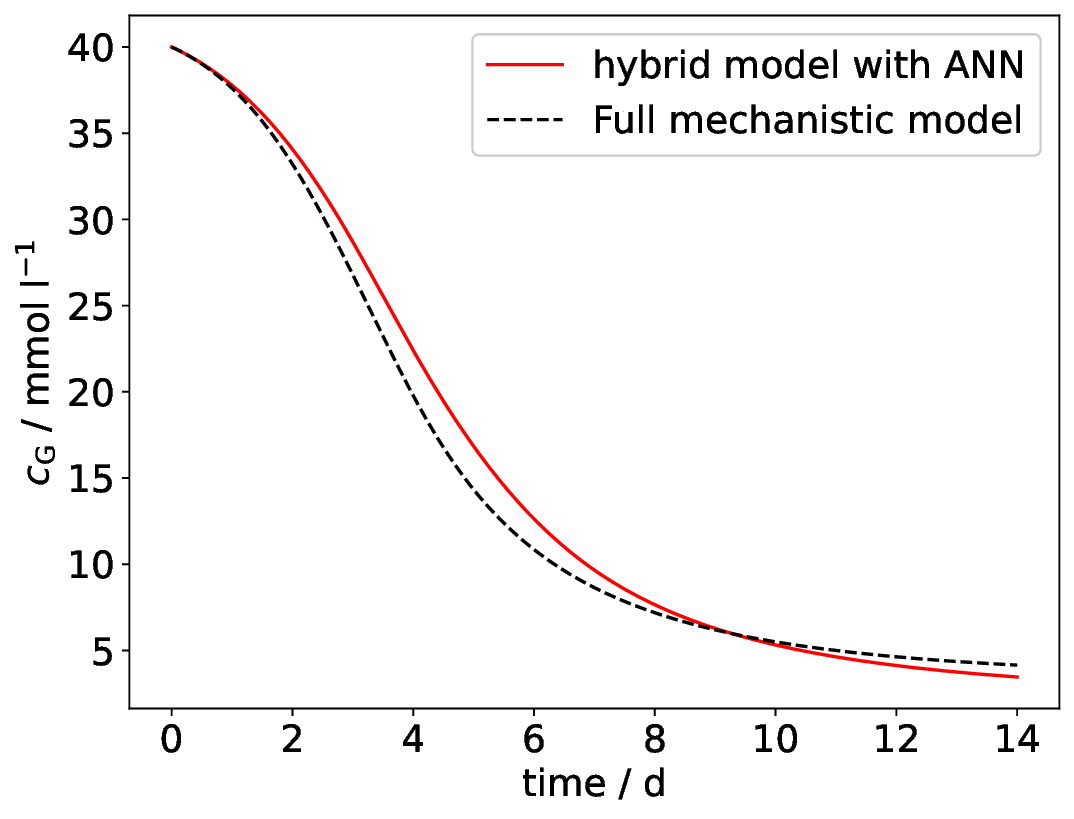}
		\caption{}
		\label{fig:bioReactor_glucose13}
	\end{subfigure}
	\caption{Simulation results with hybrid model compared to simulation results with mechanistic model for bioreactor case study. Red, solid lines: simulation results with hybrid model. Black, dashed lines: simulation results with mechanistic model. (a) Asparagine concentration. (b) Lactate concentration. (c) Cell viability. (d) Glucose concentration.}
	\label{fig:bioReactor_comparison_13}
\end{figure}

In the final step, we replace the parameters $p_1$ and $p_2$ with their corresponding data-driven models (ANNs). To analyze the performance of the resulting hybrid model, we simulate the hybrid model with an initial asparagine concentration of 13 mmol/L, which was not part of the training dataset in Step 3. We compare the hybrid model simulation results with the original bioreactor model simulation results. Figure \ref{fig:bioReactor_comparison_13} provides the simulation results, showing that the hybrid model accurately predicts the bioreactor behavior compared to the original mechanistic model.

\subsubsection{Conclusion}

The bioreactor case study demonstrates that our hybrid model identification approach delivers an accurate hybrid model, even with a very limited amount of data and a complex mechanistic part of the hybrid model.

\subsection{Research Plant with MPC}
\label{sec:caseStudies:plant}

\begin{figure}[t]
	\centering
	\includegraphics[width=0.95\textwidth]{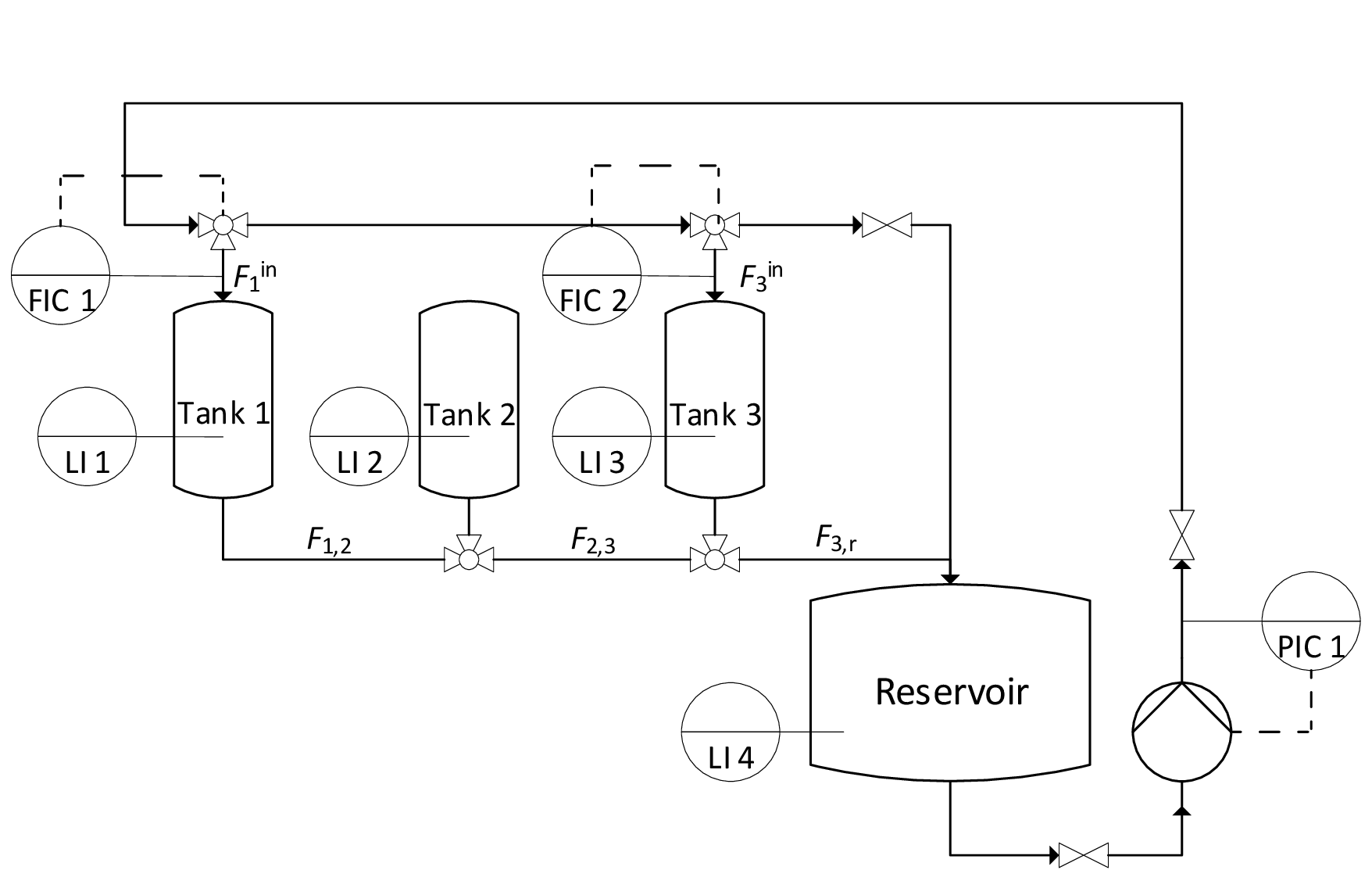}
	\caption{Flowsheet of research plant for the MPC case study. The plant is a three tank system connected to a water reservoir. The MPC manipulates the setpoints for the controllers FIC 1 and FIC 2 in order to control the tank holdup of Tank 2, measured by the level indicator LI 2.}
	\label{fig:case_study_mpc:flowsheet}
\end{figure}

\begin{figure}[t]
	\centering
	\begin{subfigure}[b]{0.45\textwidth}
		\centering
		\includegraphics[width=\textwidth]{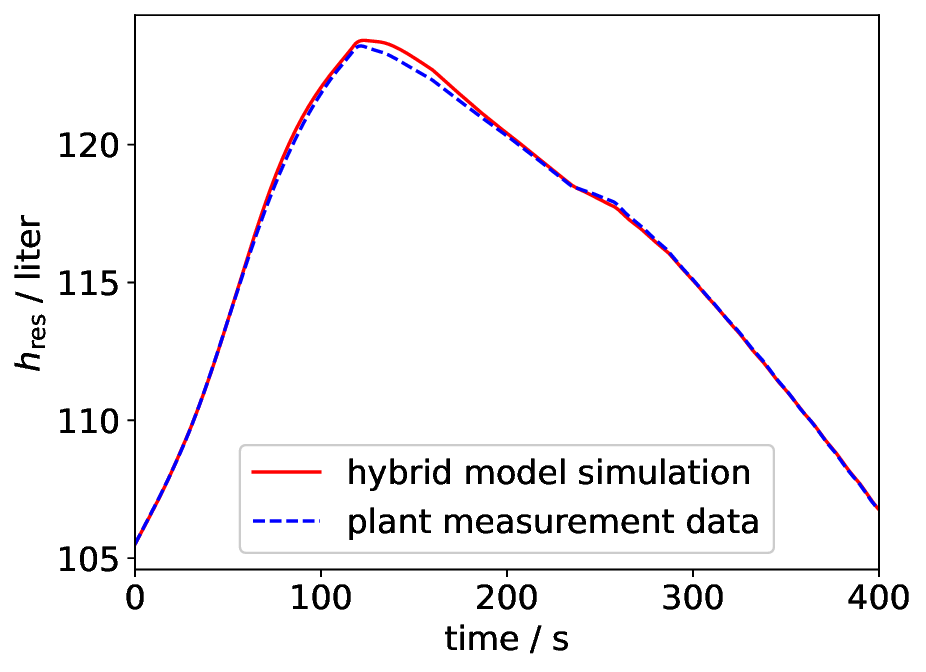}
		\caption{}
		\label{fig:sub1}
	\end{subfigure}
	\hfill
	\begin{subfigure}[b]{0.45\textwidth}
		\centering
		\includegraphics[width=\textwidth]{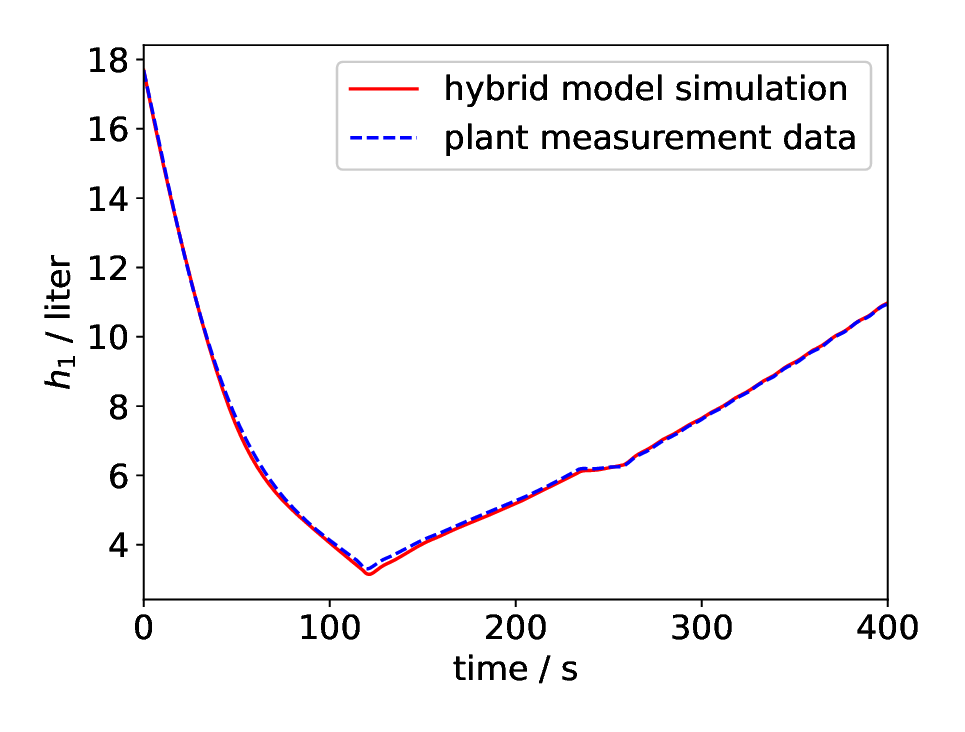}
		\caption{}
		\label{fig:sub2}
	\end{subfigure}
	\vskip\baselineskip
	\begin{subfigure}[b]{0.45\textwidth}
		\centering
		\includegraphics[width=\textwidth]{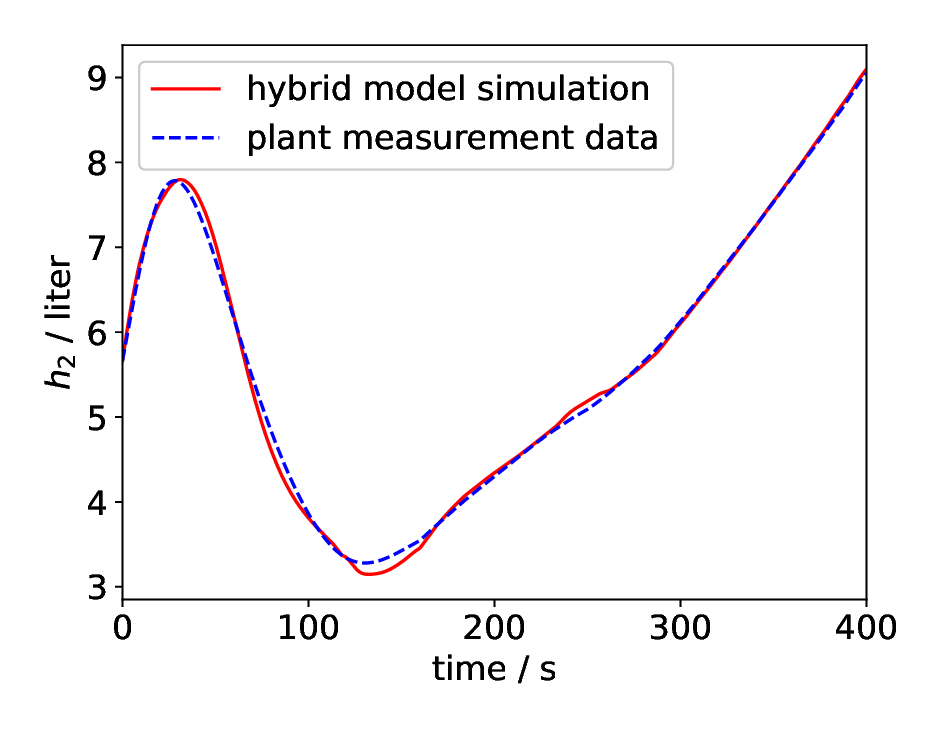}
		\caption{}
		\label{fig:sub3}
	\end{subfigure}
	\hfill
	\begin{subfigure}[b]{0.45\textwidth}
		\centering
		\includegraphics[width=\textwidth]{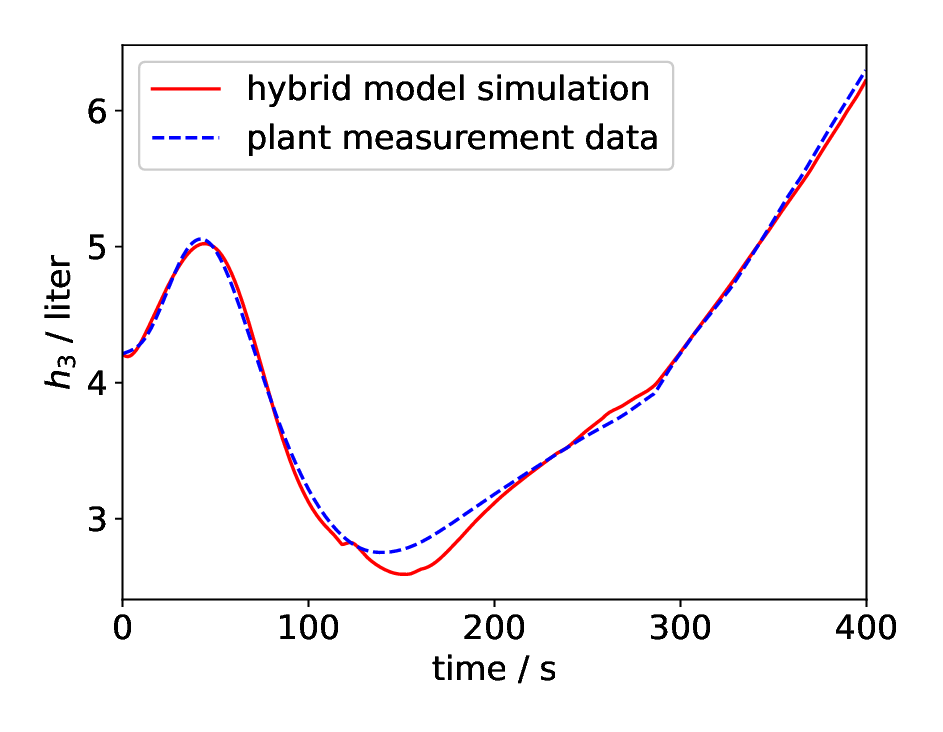}
		\caption{}
		\label{fig:sub4}
	\end{subfigure}
	\caption{Validation results of hybrid model for research plant. Plots show simulation results with hybrid model and plant measurement data. The same MV profiles are used for the hybrid model simulation and the pilot plant operation. Hybrid model simulation results are plotted with solid, red lines. Plant measurement data is plotted with blue, dashed lines. (a) Profiles for reservoir holdup. (b) Profiles for Tank 1 holdup. (c) Profiles for Tank 2 holdup. (d) Profiles for Tank 3 holdup.}
	\label{fig:caseStudies:mpc:predictions}
\end{figure}

\begin{figure}[t]
	\centering
	\includegraphics[width=0.6\linewidth]{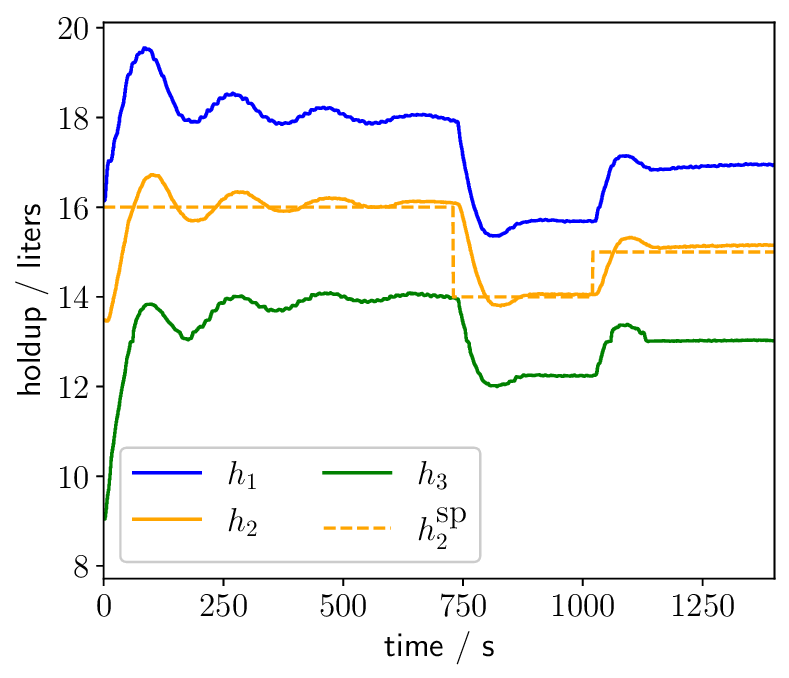}
	\caption{Results of NMPC case study for setpoint tracking.}
	\label{fig:caseStudies:mpc:closedloop}
\end{figure}

{In this case study, we identify a hybrid model for a research plant and integrate it within an MPC framework for process control.}
{As shown in Figure~\ref{fig:case_study_mpc:flowsheet}, the} plant comprises a large water reservoir and three water buffer tanks. A pump circulates the water in the pipeline, equipped with a pressure controller (PIC 1). The three buffer tanks are hydraulically connected, allowing water to flow between them and the reservoir. Tanks 1 and 3 are connected to the circulation pipeline, and their feed flow rates are controlled by flowrate controllers (FIC 1 and FIC 2). By setting flowrate setpoints for FIC 1 and FIC 2, Tanks 1 and 3 can be filled with water. Each buffer tank is equipped with level indicators (LI 1, LI 2, and LI 3), and the reservoir holdup is measured by LI 4. The flow rates between the buffer tanks and from Tank 3 to the reservoir are not directly measurable.

To control the holdups of the buffer tanks, we implement an MPC. The mass balance of the system is modeled mechanistically, while the flow rates between the buffer tanks and from Tank 3 to the reservoir are considered unknown and modeled using data-driven components. The model is formulated as follows:
\begin{subequations}
	\begin{align}
		\frac{\mathrm{d} h_1}{\mathrm{d} t}(t) & = F^\mathrm{in}_1(t) - F_{1,2}(t) \\
		\frac{\mathrm{d} h_2}{\mathrm{d} t}(t) & =  F_{1,2}(t) -  F_{2,3}(t) \\
		\frac{\mathrm{d} h_3}{\mathrm{d} t}(t) & = F^\mathrm{in}_3(t) + F_{2,3}(t)  - F_\mathrm{3,r}(t), \\
		\frac{\mathrm{d} h_\mathrm{res}}{\mathrm{d} t}(t) & = F_\mathrm{3,r}(t) - F^\mathrm{in}_1(t) -  F^\mathrm{in}_3(t) , 
	\end{align}
	\label{eq:caseStudies:model3}
\end{subequations}
with the holdups $h_1$, $h_2$, $h_3$ of Tanks 1, 2, 3, respectively, $h_\mathrm{res}$ the holdup of the water reservoir, the input flowrates $F^\mathrm{in}1$ and $F^\mathrm{in}2$ in Tank 1 and 2, respectively, and the streams $F{i,j}$ from Tank $i$ to $j$, and $F\mathrm{3,r}$ from Tank 3 to the reservoir. We consider the streams $F_{i,j}$ and $F_\mathrm{3,r}$ unknown, i.e., we model these variables with data-driven model parts using the incremental identification approach. We identify a hybrid model based on the following formulation:
\begin{subequations}
	\begin{align}
		\frac{\mathrm{d} h_1}{\mathrm{d} t}(t) & = F^\mathrm{in}_1 (t) + p_1(t) \\
		\frac{\mathrm{d} h_2}{\mathrm{d} t}(t) & =  p_2(t) \\
		\frac{\mathrm{d} h_3}{\mathrm{d} t}(t) & = F^\mathrm{in}_3 (t) + p_3(t), \\
		\frac{\mathrm{d} h_\mathrm{res}}{\mathrm{d} t}(t) & = - F^\mathrm{in}_1(t) - F^\mathrm{in}_3(t) +  p_4(t).
	\end{align}
	\label{eq:caseStudies:model3:2}
\end{subequations}
For this case study, we use implementation alternative 2 (Pyomo DAE). The MPC uses the tank holdups as controlled variables (CVs) and the setpoints for the flow rates to Tank 1 and Tank 3 (setpoints for FIC 1 and FIC 2) as manipulated variables (MVs). The MPC operates with a sampling time of 8 seconds and a control/prediction horizon of 180 seconds.

We develop a hybrid model using our incremental identification approach. To verify the hybrid model, we compare its simulation results with data from the research plant, using the same MV profiles for both. Fig.~\ref{fig:caseStudies:mpc:predictions} shows the verification results, where the hybrid model predictions closely match the plant measurement data.

We use the identified hybrid model {within the MPC} to control the research plant. Fig.~\ref{fig:caseStudies:mpc:closedloop} shows the closed-loop results, demonstrating that the MPC with the embedded hybrid model successfully tracks the desired setpoints. Note that stationary offsets can be addressed by offset-free NMPC \cite{Morari2012, Caspari2020}, which is outside the scope of this work.
% !TeX encoding = UTF-8
% !TeX spellcheck = en_US
\section{Concluding Remarks}
\label{sec:conclusions}

This work introduces a novel incremental approach for identifying dynamic hybrid models that addresses several key challenges in the field of process modeling and control. The proposed methodology comprises four systematic steps: regularized dynamic parameter estimation, correlation analysis, data-driven model identification, and hybrid model integration.

{As described in the Introduction (Section \ref{sec:introduction}) of this paper, the} above approach offers several significant advantages over traditional simultaneous identification methods. 
{It is worth noting that numerical comparisons of the two approaches have been reported in the literature \cite{Bardow2004} for particular classes of models. Our attempts to perform similar comparisons for the general dynamic hybrid systems considered in this paper were less successful, with the simultaneous approach facing significant numerical difficulties. There appeared to be two main causes for these failures:}
\begin{itemize}
	\item {\textbf{Overparameterization}: For problems with limited experimental data (cf. Case Study 2 in Section \ref{sec:caseStudies:bioreactor}), the estimation problem is vastly overparameterized because of the large number of parameters introduced by even relatively small neural networks. The incremental approach overcomes this problem via regularization, but there is no easily implementable way of achieving an equivalent effect in the simultaneous approach.}
	\item {\textbf{Numerical convergence}: as the neural network parameters have no physical meaning, the parameter estimation invariably starts from very poor initial guesses, which often results in physically unrealistic time profiles for the quantities $\boldsymbol{p}$. This problem is common to both the simultaneous and the incremental approaches. However, in the former, $\boldsymbol{p}$ is embedded within an (often complex) set of nonlinear differential and algebraic equations (DAEs); therefore, having a completely unphysical behavior for $\boldsymbol{p}$ during the early stages of the parameter estimation can create major numerical problems (e.g. with the stable integration of the system). In contrast, the incremental approach simply operates on a table of input/output data, and uses standard machine learning algorithms and tools which are well suited for dealing with bad initial parameter estimates.}
\end{itemize}

The effectiveness of our approach has been demonstrated through three diverse case studies. 
In the first case study, focusing on a continuous stirred tank reactor, we showed that complex nonlinear phenomena, including heat transfer correlations and reaction kinetics, can be accurately modeled without requiring expert knowledge or structural information of the correlations or kinetics. 
The second case study, focusing on a bioreactor, demonstrated the approach’s robustness in handling complex systems with thirty differential equations and over 2500 algebraic equations, even with limited and noisy data. 
By making use of a real research plant, the third case study validated the practical utility of our approach, successfully implementing the hybrid model within a model predictive control framework.
The successful implementation across these varied applications demonstrates the versatility and reliability of our incremental identification approach.

%As explained in the main text, this is just an "in principle" capability. 
%Most likely it would be intractable in the case of a model of any non-trivial size and/or many datasets.

Our approach offers several advantages: decoupled model development, enhanced diagnostic clarity, improved training efficiency, rigorous model evaluation, and simplified sensitivity analysis. Decoupled model development allows mechanistic and data-driven components to be developed independently, streamlining the process. Data-driven components can be used for generally unknown model parts and are not restricted to parts with a mechanistic meaning. Enhanced diagnostic clarity enables issues in mechanistic or data-driven parts to be isolated and addressed. Improved training efficiency allows standard machine learning techniques to be applied without embedding them in complex dynamic optimizations. Considering the dynamic nature of the hybrid model identification problem allows for the identification of suitable dynamic hybrid models even with very limited data. This also enables the use of data-driven models for generally unknown model parts. Rigorous model evaluation enables early assessment of the hybrid model’s structural suitability and achievable performance. Simplified sensitivity analysis allows input-output relationships in the data-driven part to be rigorously analyzed before the training of the data-driven model part takes place. This improves the prediction capabilities of the hybrid model by reducing the input space of the data-driven model parts.

{The effectiveness of the proposed approach} relies on a careful choice of the regularization parameters $\boldsymbol{W}_k^R$ (cf. Equation \eqref{eq:regularization}).
{As described in Section \ref{sec:caseStudies:chemicalReactor:regularization}, in this paper we adopted an empirical, trial-and-error approach for this task.}
Future work could focus on the development of {more} systematic approaches, something which could further enhance the method’s applicability in industrial settings. 

{
	Finally, we note that our paper assumes that the structure of the hybrid model to be identified is already chosen by the model’s developer. 
	Consequently, the methodology presented is based on a formulation (Equation~\eqref{eq:DAE}) that is general enough to accommodate a wide range of such structures. 
	One of the simplest such structures is obtained by the parameters $\boldsymbol{p}$ being added as residual terms on the right-hand sides of the differential equations (e.g. mass, energy or momentum balances). 
	This approach can work well for models involving a small number of states, such as those considered in our first and third examples (Sections \ref{sec:caseStudies:chemicalReactor} and \ref{sec:caseStudies:plant}, respectively).  
	However, many models of practical interest to process engineering applications involve hundreds or thousands of differential states. 
	In such cases, introducing an equal number of residual terms and attempting to fit a neural network which expresses them as functions of the other model variables could be problematic, particularly from the point of view of identifiability given the often limited availability of experimental data. 
	Instead, the model’s developer may wish to introduce a much smaller number of parameters to describe specific physical phenomena that are not adequately described by mechanistic equations or correlations. 
	Obviously, a degree of physical insight is required to be able to correctly identify these gaps in physical understanding. 
	The development of techniques for identifying such gaps automatically would indeed be a worthwhile topic for further research.
	Systematic approaches for model structure determination for tailored system-theoretical properties, e.g., based on semi-infinite programming as in \cite{Caspari2020}, may be relevant in this context.
}

% !TeX encoding = UTF-8
% !TeX spellcheck = en_US
\section*{Acknowledgments}
The authors acknowledge the use of Microsoft Copilot \footnote{Microsoft Copilot, \url{https://learn.microsoft.com/en-us/copilot/}, accessed: 10.08.2025}, an AI-powered writing assistant, in the drafting and refinement of this manuscript.

%\bibliographystyle{ieeetr}       % mathematics and physical sciences
%\bibliography{Literatur}   % name your BibTeX data base

\end{document}